\documentclass[journal=cmatex,manuscript=article]{achemso}

\usepackage{graphicx}
\usepackage{natbib}
\usepackage{adjustbox}
\usepackage{xcolor, hyperref}
\usepackage[version=3]{mhchem} 
\usepackage{amssymb}
\usepackage{soul}
\usepackage{color}
\usepackage{Definitions}
\usepackage{setspace}

\author{Rohit Batra}
\affiliation{School of Materials Science \& Engineering, Georgia Institute of Technology, Atlanta, GA, USA}
\alsoaffiliation{Center for Nanoscale Materials, Argonne National Laboratory, Lemont, Illinois 60439, United States}
\email{rohitbatra1989@gmail.com}

\author{Hanjun Dai}
\affiliation{Computational Science \& Engineering, Georgia Institute of Technology, Atlanta, GA, USA}

\author{Tran Doan Huan}
\affiliation{School of Materials Science \& Engineering, Georgia Institute of Technology, Atlanta, GA, USA}

\author{Lihua Chen}
\affiliation{School of Materials Science \& Engineering, Georgia Institute of Technology, Atlanta, GA, USA}

\author{Chiho Kim}
\affiliation{School of Materials Science \& Engineering, Georgia Institute of Technology, Atlanta, GA, USA}

\author{Will R. Gutekunst}
\affiliation{School of Chemistry \& Biochemistry, Georgia Institute of Technology, Atlanta, GA, USA}

\author{Le Song}
\affiliation{Computational Science \& Engineering, Georgia Institute of Technology, Atlanta, GA, USA}

\author{Rampi Ramprasad}
\affiliation{School of Materials Science \& Engineering, Georgia Institute of Technology, Atlanta, GA, USA}

\title{Polymers for Extreme Conditions Designed Using Syntax-Directed Variational Autoencoders}

\begin{document}
\newpage
\begin{abstract}
The design/discovery of new materials is highly non-trivial owing to the near-infinite possibilities of material candidates, and multiple required property/performance objectives. Thus, machine learning tools are now commonly employed to virtually screen material candidates with desired properties by learning a theoretical mapping from material-to-property space, referred to as the \emph{forward} problem. However, this approach is inefficient, and severely constrained by the candidates that human imagination can conceive. Thus, in this work on polymers, we tackle the materials discovery challenge by solving the \emph{inverse} problem: directly generating candidates that satisfy desired property/performance objectives. We utilize syntax-directed variational autoencoders (VAE) in tandem with Gaussian process regression (GPR) models to discover polymers expected to be robust under three extreme conditions: (1) high temperatures, (2) high electric field, and (3) high temperature \emph{and} high electric field, useful for critical structural, electrical and energy storage applications. This approach to learn from (and augment) human ingenuity is general, and can be extended to discover polymers with other targeted properties and performance measures.
\end{abstract}
\newpage


\section{Introduction}
A range of materials science problems are now being tackled using machine learning (ML), including discovery of materials\cite{meredig2014combinatorial,pilania2018physics,chen2019machine}, revelation of hidden structure-property relations\cite{tshitoyan2019unsupervised,sun2019data,vasudevan2019materials}, recommendations on plausible materials or chemical synthesis pathways\cite{kim2017materials}, and design of automated experiments\cite{king2009automation}. Arguably, the most direct way ML has influenced materials science is by allowing virtual screening of thousands or millions of new materials with desired functionalities\cite{kim2018polymer,xue2016accelerated}, which is far beyond the current capabilities of commonly employed high-throughput simulation or empirical techniques
This paradigm shift has resulted in the discovery of superhard ceramics\cite{mansouri2018machine}, metallic glasses\cite{ren2018accelerated}, superconducting alloys\cite{stanev2018machine}, photovoltaics\cite{lopez2017design,kanal2013efficient}, redox flow batteries\cite{cheng2015accelerating}, among others, and continues to revolutionize various aspects of materials and chemical sciences\cite{chandrasekaran2019solving,huan2017universal,pun2019physically,huan2019iterative}. 

The materials discovery process has, however, been mostly dominated by ML techniques that attempt to solve the {\it forward} problem: ``given a material, what is its property?" Under this, starting from a dataset of materials and their corresponding properties, supervised learning models are trained, using which property predictions for a large pool of plausible materials are made and candidates with desirable properties are down-selected---to be later verified by experiments. Cost-effective techniques, such as Bayesian optimization \cite{bayesianlearning}
and active learning\cite{kim2019active}, are often utilized to search over relatively larger chemical space. Nonetheless, these property-prediction approaches are restricted to (and implicitly biased towards) the initial pool of materials considered, typically falling well within the purview of human imagination. This is a major limitation considering the near infinite space offered by materials---a realistic estimate being placed around $10^{100}$, which is further compounded if one considers complications that arise due to the hierarchy of scale, stretching from sub-nanometer to microscopic and mesoscopic\cite{walsh2015inorganic}. 

Ideally, the problem of materials discovery is of {\it inverse} type: starting with a set of desired properties, the aim is to search for materials that display those functionalities.  In this regard, two distinct schemes have recently been advanced. On the one hand, global search strategies such as genetic algorithm\cite{kim2021polymer}, evolutionary searches, etc. \cite{kanal2013efficient} tackle this {\it inverse} problem by iteratively mutating/mixing material candidates until the target properties are met (or a cost function is minimized); on the other hand, deep generative ML models provide a more straightforward solution by learning an inverse functional mapping from a latent to the materials space. Although both these approaches allow us to go beyond the constraints of human imagination, the latter additionally learns inherent chemical trends present in the materials data, and allow materials design by interpolating between known suitable materials or searching within the vicinity of a known best material.

Generative models, such as variational autoencoder (VAE) \citep{gomez2018automatic, kusner2017grammar, dai2018syntax, jin2018junction, liu2018constrained} and generative adversarial network (GAN) \citep{guimaraes2017objective, you2018graph, de2018molgan}, have been recently used to design drug-like molecules \cite{kotsias2020direct, krenn2020self}, \textcolor{black}{or even polymers \cite{jorgensen2018machine} and periodic solids \cite{kim2020inverse}}.
For instance, G{\'o}mez-Bombarelli et al. \cite{gomez2018automatic} represented the chemical structure of molecules using the SMILES (or simplified molecular-input line-entry system) language, where the {\it encoder} part of the VAE learns to project the molecular SMILES onto the latent space, while the {\it decoder} unit reconstructs the molecular SMILES given its encoded latent representation. However, a key challenge in developing such a generative model for molecules (or for materials, in general) is to learn the mapping between a continuous latent space, and a discrete chemical structure space, enabled by the syntax or grammar of the SMILES language and the semantics associated with the chemistry of molecules (or materials). The absence of explicit syntax and semantics in the VAE model allows unnecessary flexibility in representing SMILES, and thus requires more training data to learn the different chemical rules underlying chemical structures (and the SMILES language). This can also results in poor quality of the learned latent space and a high occurrence of invalid SMILES upon decoding, making the process of molecular discovery inefficient.
A fundamental improvement was achieved when Kusner et al. \cite{kusner2017grammar} successfully incorporated the structure or syntax of SMILES language using the ideas of context-free-grammar. They used grammar associated with the SMILES language to convert molecular SMILES to {\it parse trees}, one-hot representation of which were then used to train the VAE models. The final missing piece of incorporating semantics (e.g., respecting elemental valency) was recently achieved by Dai et al. \cite{dai2018syntax} through the concept of stochastic lazy attributes, with the overall model termed as syntax-directed VAE. Inclusion of both syntax and semantics were shown to drastically improve the performance of the VAE by learning a more meaningful latent representation, but more importantly, it allowed decoding of---syntactically and semantically---valid molecular SMILES at a significantly higher rate, making feasible the use of generative models for materials discovery. \textcolor{black}{Recently, a very powerful and general approach, titled SELFIES, has been shown to outperform above-mentioned methods by incorporating both syntactic and semantic constraints through ‘derivation rules’ table \cite{krenn2020self}. However, extension of this method for the case of polymers, requiring special constraints, such as presence of chain ends with matching bond type, remains to be established.} The concepts of syntax and semantics are not limited to materials science, and find applications in other areas, such as symbolic expressions\cite{allamanis2017learning} and programming languages\cite{cbook1988}.

In this work, we go beyond molecular design and develop polymer-specific syntax-directed VAE to discover polymers with extreme thermal and electrical stability.
\textcolor{black}{Although VAE has been used for polymer discovery in previous works, they are either limited by the explored chemical space due to the involved grammar being restricted to a pre-determined set of chemical fragments \cite{jorgensen2018machine}, or rely on generation of large molecules as proxy to polymers \cite{st2019message, jin2020hierarchical}. In contrast, this work on polymer generation uses VAE that incorporates both syntax as well as semantics constraints.}
To achieve this, crucial modifications were introduced in the SMILES grammar, in addition to inclusion of polymer-specific semantics (e.g., 2 chain ends).
Thus, the decoding process incorporates polymer syntax and semantics using context-free-grammar parse trees and stochastic lazy attributes to ensure decoded SMILES represent plausible polymers. Another critical challenge to train a polymer VAE, as opposed to other studies on molecular VAE, is the avilability of large amounts of training examples (in the order of hundred thousands). The polymer community is particularly known to suffer from data sparsity problems. For instance, to-date the total number of chemically diverse polymers synthesized is roughly $\sim$12,000, which is not enough to learn a good-performing VAE model; in contrast, datasets consiting of over 250,000 moleucles are avilable. Here, we resolve this issue using retro-synthetic ideas to generate a representative hypothetical dataset of $\sim$250,000 polymers, constructed from a few thousand molecular `building blocks' or fragments derived from $\sim$12,000 empirically synthesized polymers obtained from the literature.

To demonstrate the preeminence of the current approach for polymer design, we use it to discover several hundred polymer candidates that are expected to perform well, and be stable, under 3 extreme conditions: (1) high temperatures, (2) high electric field, and (3) high temperature and high electric field. We make the further assumptions that high temperature and high electric field behaviors may be related to the polymer glass transition temperature ($T_{\rm g}$) and bandgap ($E_{\rm g}$), respectively. In particular, we set our property goals to be (1) $T_{\rm g}$ $>$ 600 K, (2) $E_{\rm g}$ $>$ 6.5 eV, and (3) $T_{\rm g}$ $>$ 500 K and $E_{\rm g}$ $>$ 4 eV, pertaining to the three types of extreme conditions we consider here. While high $T_{\rm g}$ polymers are desired for their mechanical robustness at elevated temperatures\cite{ho2018polymer,tan2014high,zhou2018scalable,wu2020flexible}, polymers with high $E_{\rm g}$ offer large range of electrical stability at low dielectric loss, and are useful for high-energy capacitor dielectrics\cite{mannodi2018scoping,kim2016organized}. The last design goal, namely, high temperature \emph{and} high electric field stability, is especially important for high energy density dielectrics that are both thermally stable and display high electrical breakdown strength. The occurrence of inverse-relation between the $T_{\rm g}$ and $E_{\rm g}$ of polymers makes the last goal most difficult to achieve from a materials discovery perspective.


\begin{figure}[!htb]
\begin{center}
\includegraphics[width=\textwidth]{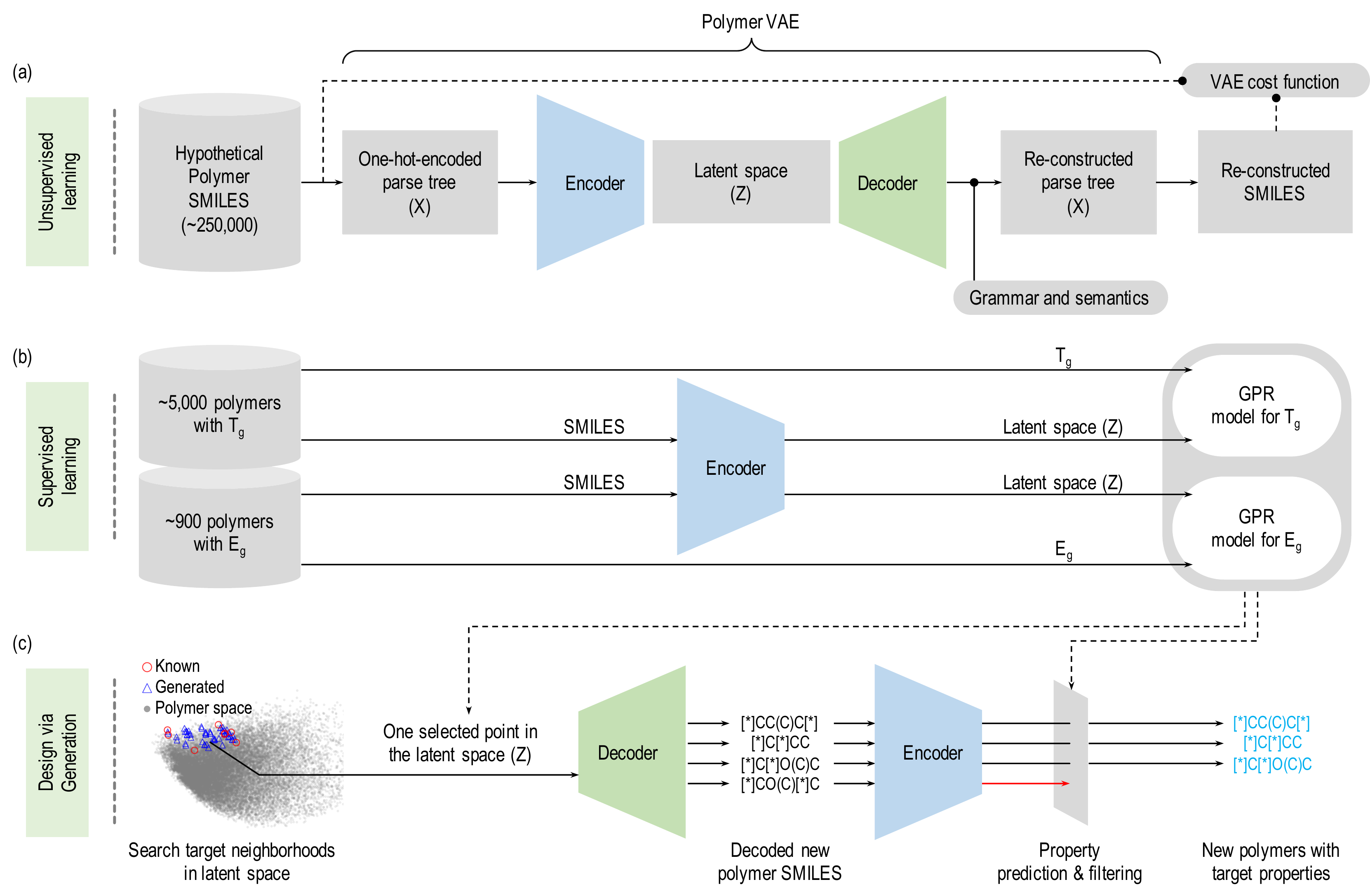}
\caption{Overall strategy to design polymers with targeted properties by solving the {\it inverse} problem of properties-to-materials space. (a) First, a variational autoencoder is used to map polymer SMILES to and from a continuous latent space. (b) Polymers with known properties ($T_{\rm g}$ and $E_{\rm g}$ here) are then encoded (or `fingerprinted' in the latent space) using the encoder. A supervised learning technique (GPR here) is then used to map these fingerprints to the different polymer properties, resulting in ML property models. (c) Finally, in the design stage, known polymers with desirable properties are first encoded to find the region of interest in the latent space. Using ML property prediction models, those latent points are sampled that satisfy a given design goal. The decoder is utilized next to construct the corresponding polymer SMILES, which go through a subsequent round of encoding and ML prediction to ensure that the decoded polymers meet the design goals.}
\label{fig:strategy}
\end{center}
\end{figure}

Figure \ref{fig:strategy} captures the overall strategy used here for polymers design. First, an unsupervised learning task is performed to train the VAE. 
Critical aspects of this process include the conversion of polymer SMILES to and from one-hot-encoded parse trees, and incorporation of grammar and semantics directly into the decoding step. The latter regulates the probability space such that only cases with valid SMILES have non-zero probability of generation. 
It should be noted that the probabilistic formulation of the latent variable in the autoencoder introduces stochasticity in the encoding and the decoding process, resulting in a distribution of polymer SMILES associated with a particular latent vector and vice-versa. 
Next, in the supervised learning stage, polymers with known properties --- $E_{\rm g}$ and $T_{\rm g}$ here---are first encoded in their latent representation, using which Gaussian process regression (GPR) models are trained for property prediction ($c.f.$ Figure \ref{fig:strategy}). Chemically meaningful latent space learned by the VAE helps build accurate GPR models for property predictions. Finally, to design new polymers with a given set of target properties, a simple enumeration followed by a generative interpolation approach  is applied.
Under this, known (or hypothetical) polymers that satisfy the given design criteria are initially encoded to find regions in the latent space where desirable polymers are expected to be present. Linear interpolations within this desired regions of the latent space are then used to select latent points for which GPR property predictions meet the desired goals. The polymer SMILES associated with such selected latent vectors are decoded, followed by another round of encoding and GPR property predictions to ascertain that the designed polymer indeed satisfies the design objectives. The full design process is illustrated under the heading of ``Design via Generation" in Figure \ref{fig:strategy}.
 
Finally, we note that the presented scheme is general, and could be used to design polymers with any set of property targets (as long as predictors for the desired set of properties are available). By establishing an {\it inverse} mapping from latent representation to polymer SMILES, it overcomes the limitations of the property-prediction (or {\it forward} problem based) approaches that screen polymers from a predetermined dataset and suffers from selection bias. Moving forward, we expect this scheme to be further refined using the concepts of transfer learning, multi-task learning and semi-supervised learning.

\section{Results and Discussion}
\subsection{Polymer Variational Autoencoder}
As stated earlier, the VAE is a type of generative model that learn to encode or decode a sequence of inputs to and from a continuous latent space. This is achieved by using deep neural networks (feed-forward, convolution or recurrent type) to represent the encoding and the decoding units, with the constraint that the encoder lowers the dimensionality of (or compresses) the input data, while the decoder performs the decompression operation to reconstruct the input sequence (see Figure \ref{fig:strategy}(a)). As an outcome of this data compression and expansion exercise, the VAE is expected to learn essential features of the data---polymer chemistry in our case. Here, we train the VAE to map the sequence of polymer SMILES to and from a continuous latent representation, allowing us to search this latent space for polymers with desired properties. Notably, VAE is an unsupervised learning method that requires just the polymer SMILES for training, without any associated property information (labeled data). Given the dearth of property data in materials science, this is a significant advantage.

Crucially, the VAE is constructed such that it incorporates both the syntax of the SMILES language and the semantics of the polymer chemistry. This is achieved using context-free-grammar parse trees and stochastic lazy attributes, details of which can be found in the section Methods. \textcolor{black}{In particular, we modified the SMILES grammar and incorporated new attributes to ensure that the generated polymers have exactly two chain ends with same bonding type (single, double bond, etc.) Other syntax and semantics constraints that are also applicable to molecules, such as valency of different elements, closure of rings, etc. were also retained (see Methods).} Henceforth, we will refer to this syntax-directed VAE as just VAE. The inclusion of grammar and semantics frees the VAE from explicitly (or spending extra efforts in)
learning such rules from the training set, and allows it to wholly concentrate on finding more important chemical trends, making the latent representation more chemically meaningful. Further, it ensures generation (or decoding) of valid polymers at a significantly higher rate, making the process of polymer design computationally efficient and practically feasible. In this work, two VAE models with latent dimensions 64 and 128, respectively termed as VAE-64 and VAE-128, \textcolor{black}{were} trained using a hypothetical polymer SMILES dataset (with $\sim$250,000 SMILES). Details on the VAE training, including the hypothetical SMILES dataset and the model architecture, are discussed in the section Methods.

\begin{figure}[t]  
\begin{center}
\includegraphics[width=\textwidth]{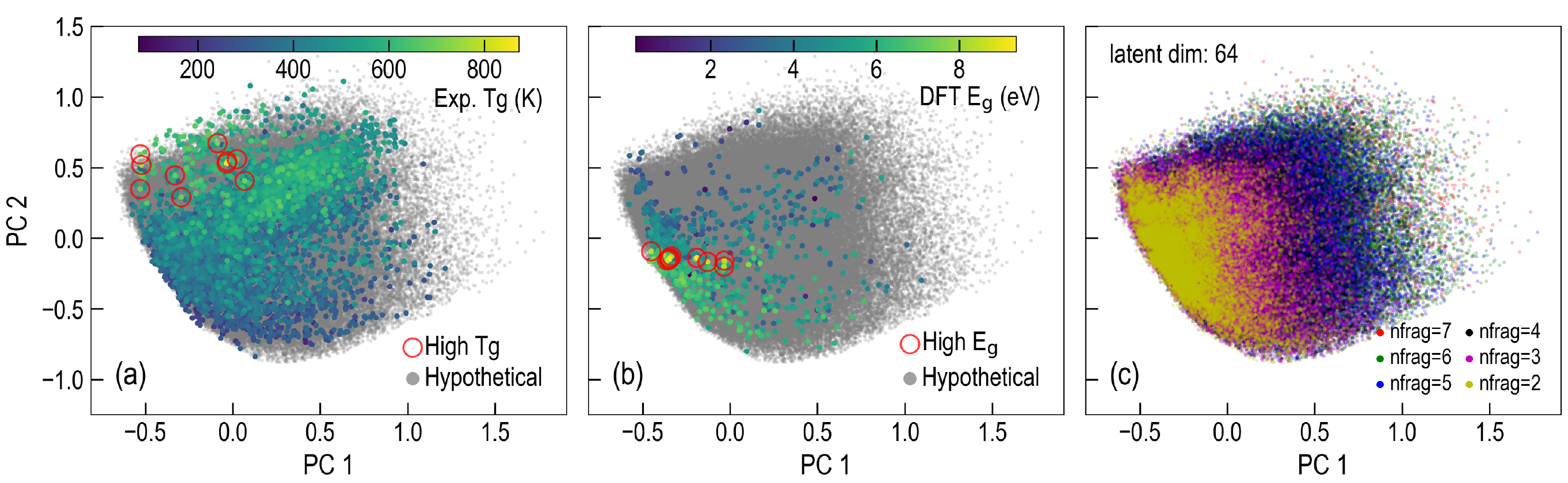}
\caption{Learned latent space of the hypothetical SMILES dataset as projected onto the first two principal components. Observed trends in the (a) empirical $T_{\rm g}$ and the (b) DFT computed $E_{\rm g}$ values when the known polymer datasets are encoded into the same latent space, with highest 10 property value polymers clustered in a limited region (red color symbols). (c) Increase in the polymer diversity with increasing `nfrag' or number of molecular building blocks (see Methods) as reflected by the area spanned in the latent space.}
\label{fig:latent_space}
\end{center}
\end{figure}

Figure \ref{fig:latent_space} captures the ability of the polymer VAE to learn a chemically meaningful latent representation through just the unsupervised learning of SMILES. In panels (a) and (b), gray colored symbols mark the latent projections of the entire hypothetical SMILES dataset, reduced into the first two principal components. Likewise, the polymers with known $T_{\rm g}$ and $E_{\rm g}$, respectively named as $T_{\rm g}$ and $E_{\rm g}$ datasets, are converted to their latent representations, followed by their projection onto the same two principal components. More details on these two property datasets, including their source and data size, can be found in the section Methods. Clear trends on the concentration of polymers with similar $T_{\rm g}$ and $E_{\rm g}$ in Figure \ref{fig:latent_space}(a) and (b) suggest that the VAE has indeed learned important chemical characteristics. Moreover, red colored symbols denoting the top 10 known polymers with highest $T_{\rm g}$ and $E_{\rm g}$ values appear in a small region of the latent space, highlighting the area where similar polymers with perhaps even higher $T_{\rm g}$ and $E_{\rm g}$ values can be found. The presence of these chemical trends can be explained based on the following rationale: during training, VAE learns to place polymers with similar SMILES in the neigboring regions of the latent space, thereby automatically learning the chemical information present in the SMILES representation. We contrast this observation with the previous study on molecules, where G{\'o}mez-Bombarelli {\it et al.} \cite{gomez2018automatic} were unable to see such clear chemical trends. Possibly, this is because their VAE model does not respect SMILES grammar and molecular semantics, as was discussed in the recent works\cite{kusner2017grammar}. Lastly, in panel (c) we demonstrate how the diversity of the hypothetical SMILES dataset increases with the increase in the number of molecular building blocks chosen to create the polymers. While Figure \ref{fig:latent_space} presents results for VAE-64, similar observations using VAE-128 are provided in Supporting Information (SI) Figure S1.
 
Other metrics to measure the performance a VAE are the reconstruction accuracy and the prior validity. The former captures the ability of the model to reconstruct polymer SMILES from their corresponding latent representations, a measure of the quality of the latent space; the latter measures the probability of the decoder to generate valid SMILES when sampling from the prior latent space and quantifies the computational efficiency of the decoding process. Considering the probabilistic nature of the encoder and the decoder, the reconstruction accuracy was computed by encoding each polymer 10 times, followed by decoding 100 times. This results in 1000 decoded SMILES for each input polymer. Two types of reconstruction accuracy were defined: 1) strict, when the most frequent of these 1000 decodings were same as the input polymer, and 2) loose, when at least one of the 1000 decodings were same as the input polymer. The prior validity was computed by sampling 1000 latent points from the prior distribution $p(\mathbf{z}) = \mathcal{N}(0,\mathbf{I})$, decoding each of these points 500 times, and finding the overall percentage of valid SMILES generated. Good performance for the average reconstruction accuracy for the $T_{\rm g}$ and the $E_{\rm g}$ datasets, and the percentage prior validity for the case of VAE-64 and VAE-128 can be seen in Table \ref{tab:vae_performance}. We note that both $T_{\rm g}$ and $E_{\rm g}$ datasets were not included during the VAE training, and constitute unseen `new' polymer cases. While the reconstruction accuracy of VAE-128 is better than VAE-64, perhaps owing to larger latent dimensionality or capacity to retain information, its performance is slightly worse in terms of prior validity.
A fair comparison between the VAE results presented here is not possible with the past works on molecules owing to the different dataset types and variations in chemical diversity. However, we note that similar level of performance was seen in the past for the case of molecules; reconstruction accuracy and prior validity were respectively reported to be 44.6 and 0.7 without incorporating grammar and semantics\cite{gomez2018automatic}, 53.7 and 7.2 with grammar only\cite{kusner2017grammar}, and 76.2 and 43.5 with both grammar and semantics\cite{dai2018syntax}.

\begin{table}[h]
\caption{Performance of the polymer VAE as a function of latent dimensions for the $T_{\rm g}$ and $E_{\rm g}$ datasets. See text for the details on the definitions of two types of reconstruction accuracy, strict and loose (in brackets), and prior validity. $R^2$ coefficient and RMSE on the test set (20 \% of dataset) for the two ML (GPR) models are also provided.}
\label{tab:vae_performance}
\centering
\begin{tabular}{c c c c c c}
\hline
\hline
Model & Prior validity (\%) 	& \multicolumn{2}{c}{Reconstruction (\%)}	& \multicolumn{2}{c}{ML $R^2$ (RMSE)} \\\cline{3-6}
&  	& $T_{\rm g}$ (K)&	 $E_{\rm g}$ (eV)& $T_{\rm g}$ (K)	&	$E_{\rm g}$ (eV)\\
\hline
VAE-64	   &	 27     & 51 (75)	&	 53 (86) 	& 0.76 (51.63)	&	 0.50 (1.04) \\
VAE-128   &	 13		& 57 (78)	&	 67 (89) 	& 0.79 (48.32)	&	 0.53 (1.03) \\
\hline
\end{tabular}
\end{table}

\begin{figure}[ht]  
\begin{center}
\includegraphics[width=0.8\textwidth]{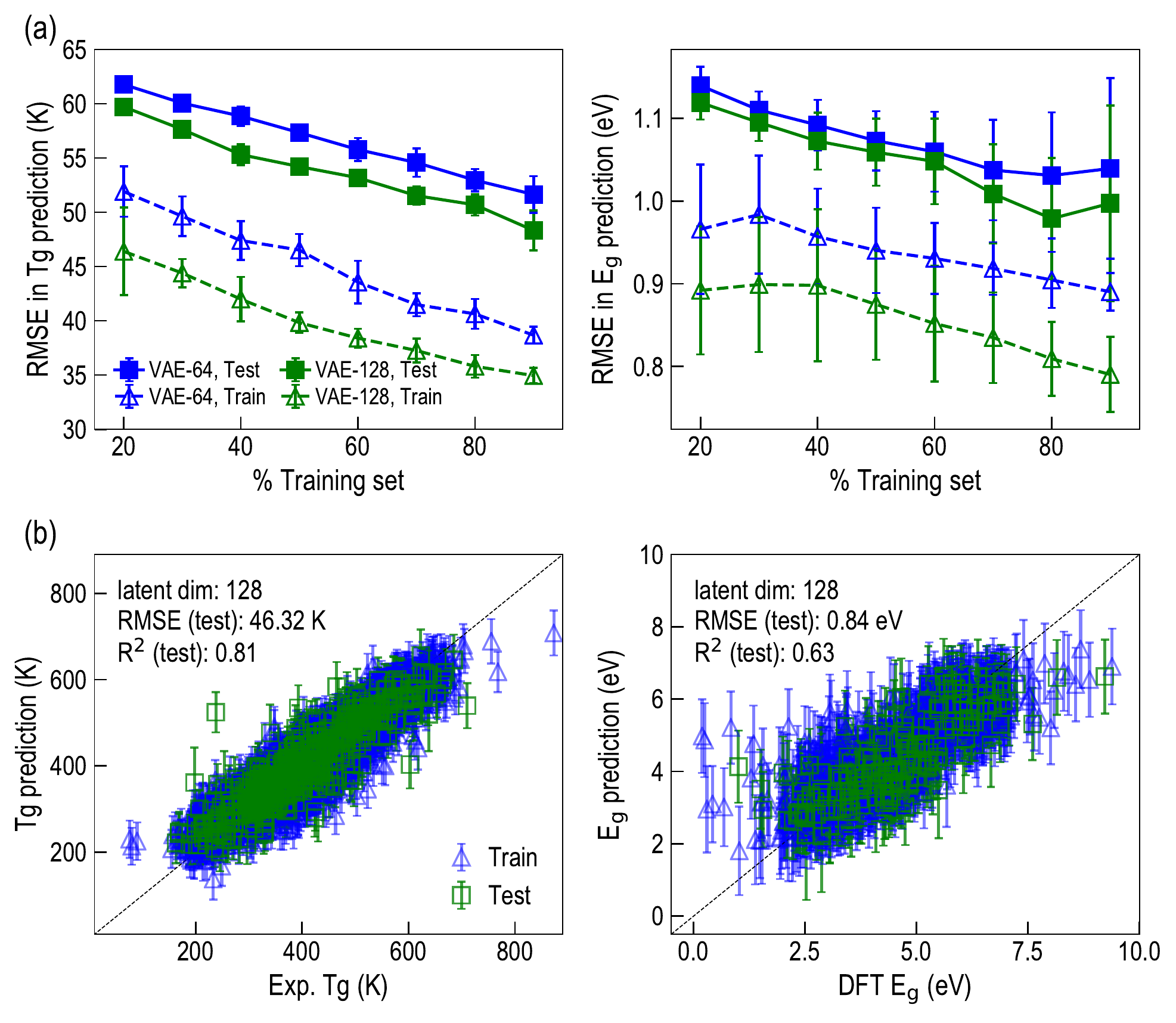}
\caption{(a) Learning curves for the GPR models built using the $T_{\rm g}$ and $E_{\rm g}$ datasets. The polymers were first encoded using VAE, which was then mapped to their corresponding $T_{\rm g}$ or $E_{\rm g}$ property values. (b) Example parity plots demonstrating the performance of the GPR $T_{\rm g}$ and $E_{\rm g}$ models built using VAE-128 latent representation.}
\label{fig:lc}
\end{center}
\end{figure}
Having established the validity of the latent space learned by polymer VAE, we next build GPR models using the $T_{\rm g}$ and $E_{\rm g}$ datasets. For this, the polymer latent representations obtained from the VAE model were mapped to their respective property values (see Methods). 
The learning curves for the two properties using both the VAE-64 and VAE-128 latent representations are provided in Figure \ref{fig:lc}(a). In all cases, reasonable error trends with the test error decreasing with increasing training set size can be seen. Provided $T_{\rm g}$ dataset is estimated to have a noise of around 30 K, reaching test errors around 50 K is quite respectable. 
The performance of $E_{\rm g}$ models, although acceptable, is comparatively worse. This could be because of the larger inherent noise present in the theoretically computed $E_{\rm g}$ values of polymers, especially with the assumption of treating them as single chains (see details on source of $E_{\rm g}$ dataset in Methods section). In corroboration with the results obtained for the reconstruction accuracy, the performance of VAE-128 GPR models is visibly better than VAE-64 for both the property datasets. Example parity plots illustrating the performance of the GPR property prediction models, using VAE-128 latent representations, are shown in Figure \ref{fig:lc}(b). Two quantitative error measures, i.e., root mean square error and correlation coefficient, averaged over 10 statistical runs, are also provided in Table \ref{tab:vae_performance} for both the datasets. Once again, we note that the good level of accuracy reached by the GPR models is a direct indicator of the high quality of latent representation learned by the VAE.

\subsection{Polymer Design}
As example problems, we choose to design three \textcolor{black}{classes} of polymers with the following property objectives: (1) $T_{\rm g}>600$ K, (2) $E_{\rm g} > 6.5$ eV and (3) $T_{\rm g} > 500$ K and $E_{\rm g} > 4$ eV. Polymers with these functionalities are desirable for various dielectric applications; high $T_{\rm g}$ polymers provide high mechanical strength over larger temperature window \cite{wu2020flexible}, high $E_{\rm g}$ polymers intrinsically have low dielectric loss and high enough electrical breakdown strength \cite{kamal2020computable}, and high $T_{\rm g}$ and $E_{\rm g}$ polymers are expected to be great candidates for high energy density capacitor applications, particularly, at higher temperatures. \textcolor{black}{While at low temperatures bandgap has been found to correlate well with the breakdown strength of polymers \cite{kamal2020computable}, at high temperatures other factors, such as chain-packing or defects may become more relevant. Thus, high $E_{\rm g}$ as a criterion for design of high breakdown strength polymers is only approximate, and can be replaced with more accessible or correlated properties.} The difficulty level of these three design goals becomes apparent from the $T_{\rm g}$ vs $E_{\rm g}$ plot in Figure \ref{fig:egtg}(a), wherein all the 303 known polymers common to both the property datasets are drawn. The number of polymers known to satisfy the aforementioned 3 design goals, i.e., high $T_{\rm g}$, high $E_{\rm g}$, and high $T_{\rm g}$ and $E_{\rm g}$ are 9, 59 and 17, respectively. Further, an inverse relation between the $T_{\rm g}$ and $E_{\rm g}$ \textcolor{black}{illustrates} the difficulty of the last design goal. Finally, we note that our aim here is to find chemically diverse polymers that satisfy these design goals rather than finding polymers that maximize/minimize a particular property. This is relevant from an application standpoint because for a successful dielectric, a polymer must satisfy a collection of properties, including high charge transport barrier, high breakdown strength, low dielectric loss, etc., rather than display one best property. By imposing many such simple design criteria the hope is that several polymer candidates can be proposed in extensions to this work, which can be later down-selected by imposing other screening criteria based on a target application.

\subsubsection{Design via Enumeration}
\begin{figure}[ht]  
\begin{center}
		\includegraphics[width=\textwidth]{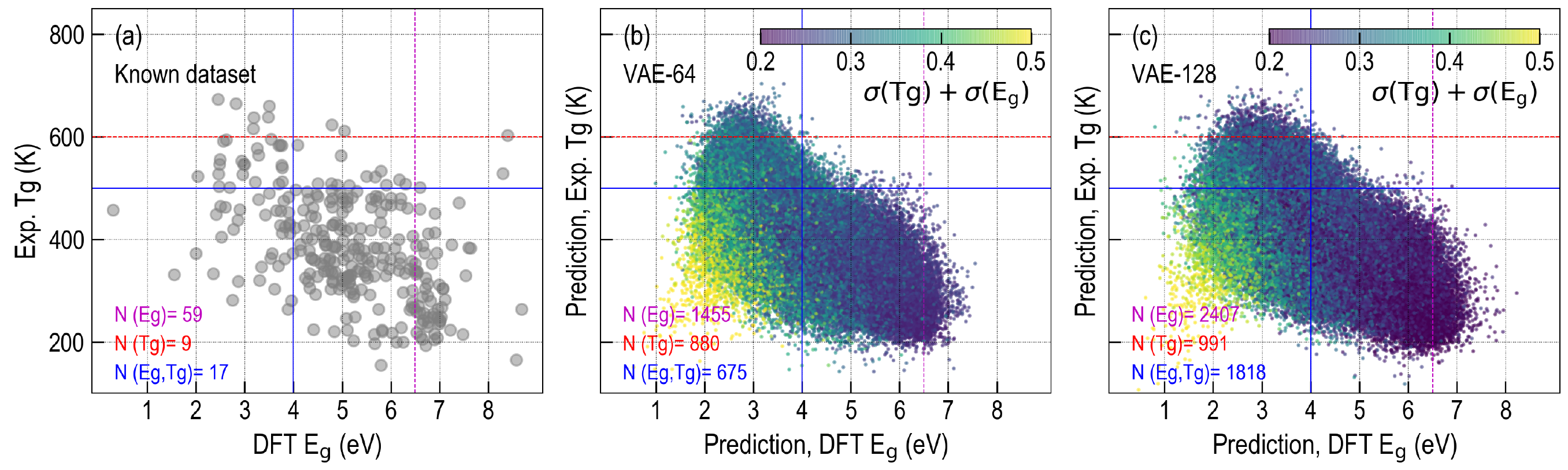}
		\caption{Comparison of the trend between $T_{\rm g}$ and $E_{\rm g}$ for (a) known polymers, and for hypothetical polymers as predicted using GPR models based on latent representations of (b)VAE-64 and (c)VAE-128. In (b) and (c), the marker color represent the sum of the uncertainties in GPR $T_{\rm g}$ and $E_{\rm g}$ predictions, each normalized by its range. The magenta, red and blue colored lines are guide to the three design criteria high $E_{\rm g}$, high $T_{\rm g}$, and high $E_{\rm g}$ and $T_{\rm g}$, respectively, with N($E_{\rm g}$), N($T_{\rm g}$) and N($E_{\rm g}, T_{\rm g}$) highlighting the number of polymers found to satisfy the three design criteria (in same order) in different datasets.}
		\label{fig:egtg}
	\end{center}
\end{figure}
As part of the first design problem, we simply make property predictions for the entire hypothetical SMILES dataset of nearly 250,000 polymers. The $T_{\rm g}$ and $E_{\rm g}$ GPR models were used to make these predictions based on the VAE encoded latent representations. The inverse relation between the $T_{\rm g}$ and $E_{\rm g}$ as observed for the case of known polymers in Figure \ref{fig:egtg}(a), is also preserved when using the GPR property predictions for the hypothetical polymer dataset in Figure \ref{fig:egtg}(b) and (c). This, along with test set accuracy reached by the GPR models, provides some confidence in the property predictions made for the hypothetical polymer dataset using the present scheme. There are, however, some polymers with high uncertainty in the GPR predictions, as captured by the marker color codings in panels (b) and (c). Nonetheless, the yellow colored high uncertainty regions are different from those where the three design criteria are met. The number of hypothetical polymers that satisfy the three design criteria are also included in panels (b) and (c), which is significantly larger than that in the original $T_{\rm g}$ and $E_{\rm g}$ datasets. The complete list of these new polymer candidates predicted to have the desired functionalities, and designed through solving the {\it forward} problem is provided in SI.

\subsubsection{Design via Inversion}
The chemical trends captured by the latent representation, illustrated in Figure \ref{fig:latent_space}, suggest that polymers with similar properties occupy closer neighborhood in the latent space. Thus, if a few polymers with the desired properties are already known, a search in their latent neighborhood could reveal many more suitable polymer candidates. The GPR property prediction models could be used to advance this search. Once the latent points satisfying the given design criteria are identified, the {\it inverse} mapping from the latent representation to the polymer, made available through the VAE decoder, can be used to find the associated polymer candidate. We note that owing to the stochastic nature of the decoding process, there is some ambiguity with the exact polymer represented by the selected latent point. Thus, to address this issue, a second round of encoding for all the decoded polymer candidates is performed and the associated GPR property predictions are made to check if the design criteria are indeed met. Polymers satisfying the design criteria are retained.
\begin{figure}
	\begin{center}
		\includegraphics[width=0.71\textwidth]{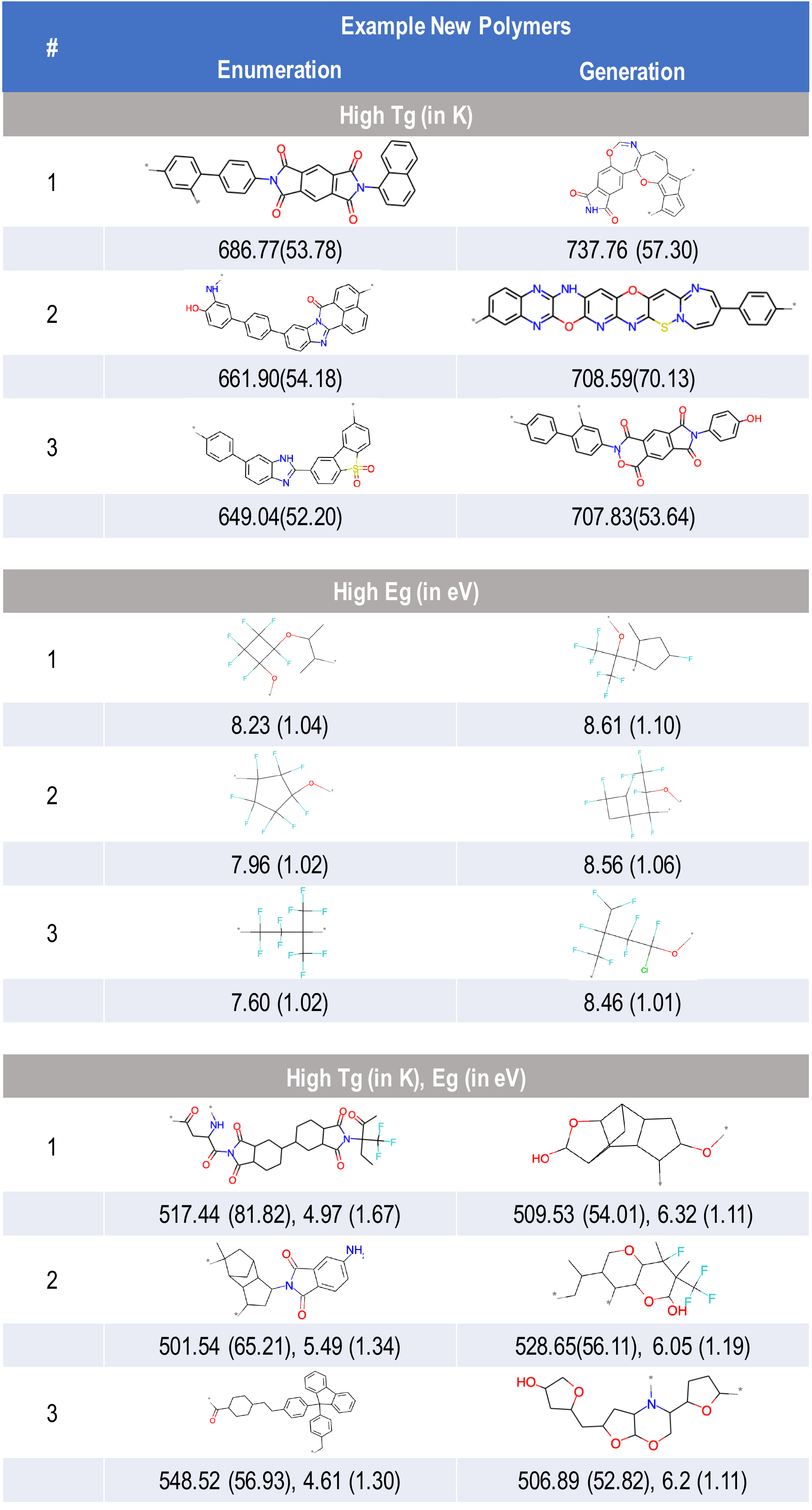}
		\caption{Example polymers designed in this work, along with their GPR Tg and E$_{\rm g}$ estimates and uncertainties (in brackets). The symbol `*' represents the polymer chain ends.}
		\label{fig:candidates}
	\end{center}
\end{figure}

Following the above-mentioned design strategy, for a particular design goal, the top 250 polymer candidates identified as part of the earlier enumeration scheme were selected. Linear interpolations in the latent space for all combinations of selected 250 polymers were performed, and latent points whose associated GPR predictions meet the design criteria were selected. Owing to computational cost, decoding for only top 10,000 latent points---with the highest GPR property prediction values---was done 100 times. Only those decoded candidates were retained that satisfy the design criteria after another round of encoding, as discussed above and illustrated in Figure \ref{fig:strategy}(c). This procedure was repeated for all three design criteria. The complete list of polymer candidates generated using this approach are provided in SI, while some representative cases are presented in Figure \ref{fig:candidates}. Care was taken to eliminate duplicate candidates with the $E_{\rm g}$, $T_{\rm g}$ and the hypothetical SMILES datasets.

A few observations can be made from the results presented in Figure \ref{fig:candidates}. First, from a chemical standpoint the proposed polymer candidates appear to be rational; predominance of conjugated systems is reflective of polymers with high $T_{\rm g}$, presence of saturated C-F bonds suggest high $E_{\rm g}$, and a balance between saturated rings and C bonded with highly electronegative F and O species is expected to have both high $T_{\rm g}$ and  $E_{\rm g}$. Though synthesis of these candidates, and their subsequent characterization and measurement would be required to confirm their property predictions. Nonetheless, for the case $E_{\rm g}$ predictions we provide preliminary validation using DFT computations. We selected 30 new polymers (10 from each of the design goals) proposed using VAE design scheme, and constructed single chain polymer systems on which DFT calculations were performed, consistent with generation of the $E_{\rm g}$ training dataset. The results are similar to those observed for test set in Figure \ref{fig:lc} and are provided in SI Figure S2.

Second, following the simple strategy of search in the latent space neighborhood, the generative or {\it inverse} scheme is indeed able to identify several cases which have superior properties (higher $T_{\rm g}$ and/or $E_{\rm g}$) than that obtained through enumeration or property-prediction ({\it forward} problem) approach (also see SI). This clearly highlights the advantage of the design strategy presented here (which allows us to go beyond the constraints of human imagination) and the motivation to solve the {\it inverse} problem. \textcolor{black}{In contrast to the 17 known polymers, almost 300 new potential candidates have been designed (list provided in SI) with the goal of exhibiting high $T_{\rm g}$ and $E_{\rm g}$. Further, this scheme has the potential to find many more polymers satisfying these criteria.} Lastly, associated with each distinct design goal, polymers with very different chemistry are generated by the decoder. An analysis of the common chemical fragments present in the VAE generated polymers for the three target properties is provided in SI. \textcolor{black}{It not only shows the distinct chemistry necessary to satisfy each of the three conditions, but also reveals new trends in form of chemical fragments promoting these properties. For instance, saturated 4,5 member rings, bridged rings, oxolane groups, and C atoms bonded to a limited number of electronegative F, Cl and O groups had higher occurrence rate for polymers predicted to have high $T_{\rm g}$ and $E_{\rm g}$.} Furthermore, since the VAE generated polymers contain `new' chemical fragments as opposed to those present in the training data, it demonstrates the ability of our approach to explore extremely huge chemical space, without being restricted to a pre-determined set of chemical fragments. Overall, the VAE approach resulted in generation of resonable polymers, with distinct chemistry for different conditions, involving generation of `new' chemical fragments and with superior property predictions. We expect the inclusion of grammar and semantics in the VAE as the major cause that enabled learning of a meaningful latent space, making it a powerful polymer-discovery tool.

Many avenues remain for improving the design strategy adopted here. The set of polymer semantics included is not complete. Incorporating more semantics is expected to enhance the performance of the entire design process. \textcolor{black}{The model provides no information regarding the synthesizability of the generated polymers; this point is particularly relevant for high $T_{\rm g}$ polymers that are difficult to synthesize.} Moreover, rather than individually conducting the unsupervised (autoencoder) and the supervised (GPR) training, joint training can be performed for all the SMILES as well as the property datasets. Such semi-supervised learning has been shown to improve the performance of machine learning models. \textcolor{black}{Efforts to incorporate other aspects of polymer structure, such as crosslinking, chain length, dispersity, etc. should also be made, perhaps through the use of recently developed BigSMILES representation \cite{lin2019bigsmiles}. Further, polymer search in the latent space could be potentially improved using reinforcement learning \cite{zhou2019optimization}.}

\section{Conclusion}
The space of potential materials is near-infinite, making the search for materials with desirable properties intractable. With the introduction of machine learning, the design of new materials with desirable properties has accelerated and/or improved. However, still most of the methods rely on solving the {\it forward} problem, wherein materials-to-property mappings are learned using which a predetermined list of material candidates are screened. This approach is not only limited in terms of the explored materials space, but is computationally inefficient. Here, instead we demonstrate a strategy that solves the {\it inverse} problem: starting from the properties required, desirable materials candidates are generated. As an example, we design 3 classes of polymers applicable for thermal extremes, electrical extremes, and thermal and electrical extremes.

Our strategy is based on the use of variational autoencoders
to encode  and decode polymers to and from a continuous latent space, allowing one to search this space for new polymers with desirable properties. The autoencoder is trained to operate using the SMILES representation of the polymers. Crucially, the learning ability of the autoencoder is enhanced by directly incorporating the SMILES grammar and polymer semantics using context-free-grammar parse trees and stochastic lazy attributes. This allows to construct a more meaningful latent representation for polymers, and improves the efficiency of the decoder to generate valid polymers. Next, supervised learning GPR models were built for two property datasets, i.e., $T_{\rm g}$ and $E_{\rm g}$, using the polymer latent representations.

To design polymers with desired $T_{\rm g}$ and $E_{\rm g}$ properties, enumeration and generative schemes were adopted. In the former case, GPR property predictions for a hypothetical (but realistic) polymer dataset were made to screen polymers that meet the design criteria, parallel to solving the {\it forward} problem. These selected polymer candidates helped drive the latter case by identifying interesting latent neighborhoods, which were searched using GPR property prediction models to find latent points that correspond to the required properties. The decoder was then used to find the associated polymers. More importantly, among the two approaches, the generative design approach (which solves the {\it inverse} problem) was found to discover polymer candidates with superior properties, with a much lower computational budget. Using both these schemes, we proposed several hundred potential polymers for different dielectric applications, although we note that the proposed design scheme is quite general and can be applied to discover polymers with any set of desired functionalities. Further, we expect the advantage of the generative design to become more apparent when it is used to perform targeted search in the latent space for polymers with more complex set of properties.

\section{Methods}
\subsection{Unsupervised Learning: Polymer Variational Autoencoder}

A variational autoencoder\cite{kingma2014auto} is a framework to learn the generative model with latent variables for the input dataset $\mathbf{X} = \left[\mathbf{x}_1, \mathbf{x}_2,..., \mathbf{x}_N\right]$. The generative process of each sample $\mathbf{x}$ is performed by first sampling a latent vector $\mathbf{z}$ from prior $p(\mathbf{z})$ (typically a normal Gaussian distribution, denoted as $\mathcal{N}(0, I)$), and then sample from conditional distribution $p_{\theta}(\mathbf{x} | \mathbf{z})$ (known as `decoder'). The overall goal is to maximize the marginal likelihood $p(\mathbf{x})$, but due to the intractability of the direct optimization, a variational posterior $p_{\phi}(\mathbf{z}|\mathbf{x})$ (known as `encoder') is introduced to optimize the evidence lower bound (ELBO): 
\begin{eqnarray}
	\mathcal{L}(\phi,\theta; \mathbf{x}) = & \log \int_{z} p_{\theta}(\xb | \zb) p(\zb) \nonumber \\ 
	& \geq \EE_{q_{\phi}(\zb|\xb)} \big[ \log p_{\theta}(\xb|\zb) + \log p(\zb) - \log q_{\phi}(\zb|\xb) \big]
\end{eqnarray}
To make it easy to sample or calculate the probability from $q_{\phi}(\zb|\xb)$, typically a parametric distribution is assumed (in this work, we assume a multivariate gaussian with diagonal covariance matrix). 
By limiting the dimensionality of $\zb$, the encoder is forced to learn meaningful representations in latent space. With a carefully designed decoder, the decoding probability mass will be reshaped towards desired target region. 
Next, we introduce two important concepts of context-free-grammars and stochastically lazy attributes which enforce the decoder to (re)construct syntactically and semantically valid polymer SMILES.

\subsubsection{Incorporating Grammar or Syntax}
\begin{figure}[!htb]  
	\begin{center}
		\includegraphics[width=0.65\textwidth]{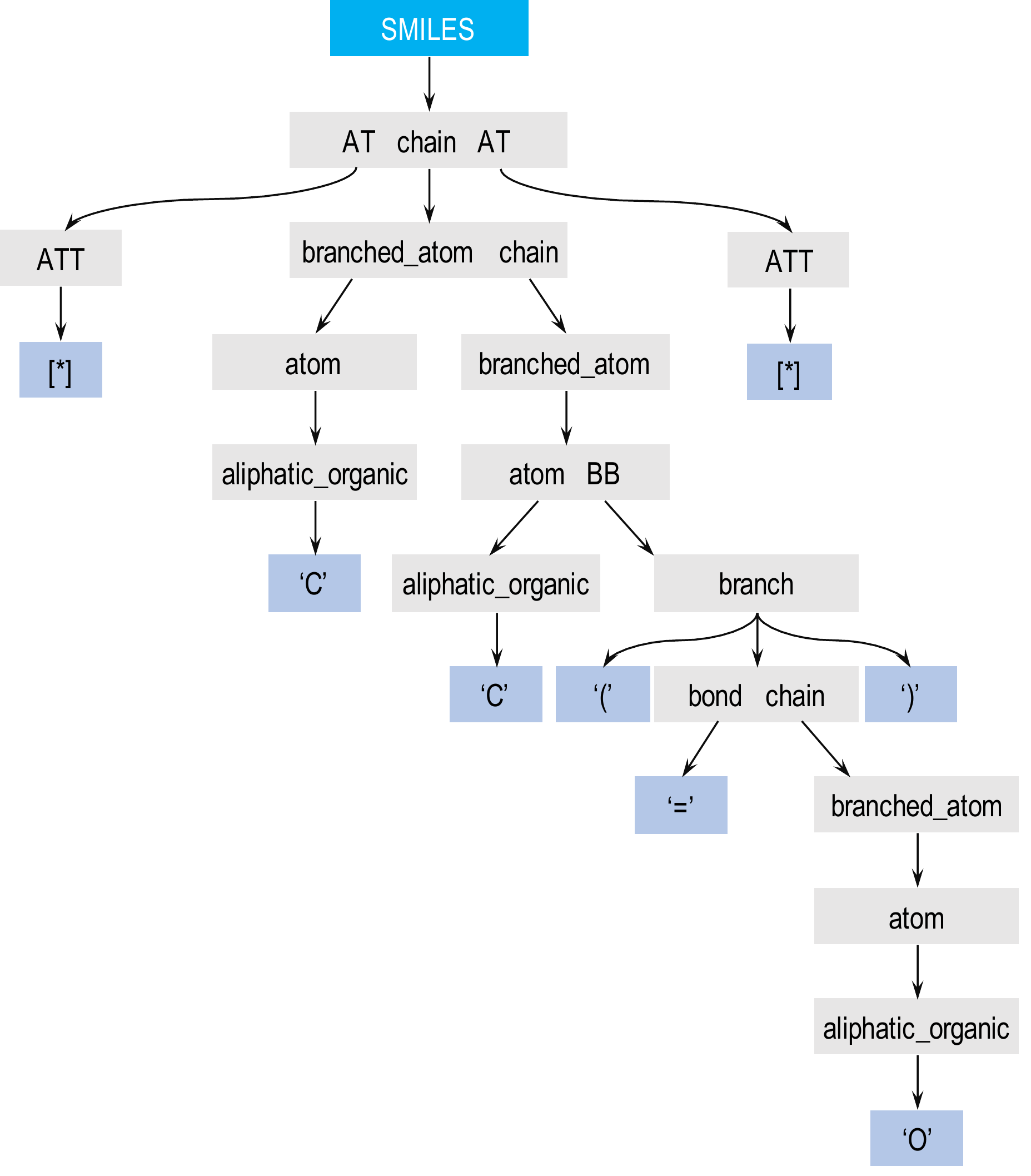}
		\caption{Syntactic tree for polymer SMILES [*]CC(=O)[*] denoting a polyketone. The tree is generated by recursively applying production rules present in the grammar (see text for details), until all leaf nodes contain terminal symbols.}
		\label{fig:cfg}
	\end{center}
\end{figure}

Given a language and its context-free-grammar (CFG), a sentence in this language ($e.g.$ a SMILES string) can be transformed to a {\it parse tree}\cite{grammarbook2006}. While this concept is common for processing sentences in natural or programming languages, here we utilize it to transform polymer SMILES into its syntactic tree structure, which will be used as an input/output to the VAE. Formally, a CFG is defined as $G = (V, \Sigma, R, S)$, where $V$ is a finite set of non-terminal symbols; $\Sigma$ is a finite set of terminal symbols, distinct from V; $R$ is a finite set of production rules; and $S \in V$ is a distinct non-terminal start symbol. Each production rule $r \in R$ is denoted as $r = \alpha \rightarrow \beta$ for $\alpha \in V$ is a nonterminal symbol, and $\beta = u_1u_2 . . . u_{|\beta|} \in (V \cup \Sigma)^*$ is a sequence of terminal and/or nonterminal symbols. Thus, for a string in a language---i.e. a sequence of terminals in $\Sigma$---the grammar G converts it into a syntactic tree, consisting of branching of non-terminal symbols in $V$, produced by recursively applying rules in $R$ to leaf nodes, until all leaf nodes are terminal symbols in $\Sigma$. For example, a parse tree for a polyketone, with SMILES [*]CC(=O)[*], is shown in Figure \ref{fig:cfg}; the `[*]' symbolizes the chain ends. This was generated using the grammar G provided below:\\
\begin{spacing}{1.0}
{\small
\noindent smiles $\rightarrow$ chain $|$ AT chain AT $|$ AT chain\\
AT $\rightarrow$ ATT $|$ ATT bond $|$ bond ATT\\
ATT $\rightarrow$ `[*]'\\
atom $\rightarrow$ bracket\_atom $|$ aliphatic\_organic $|$ aromatic\_organic\\
aliphatic\_organic $\rightarrow$ `B' $|$ `C' $|$ `N' $|$ `O' $|$ `S' $|$ `P' $|$ `F' $|$ `I' $|$ `Cl' $|$ `Br' $|$ `H'\\
aromatic\_organic $\rightarrow$ `c' $|$ `n' $|$ `o' $|$ `s'\\
bracket\_atom $\rightarrow$ `[' BAI `]'\\
BAI $\rightarrow$ isotope symbol BAC $|$ symbol BAC $|$ isotope symbol $|$ symbol\\
BAC $\rightarrow$ chiral BAH $|$ BAH $|$ chiral\\
BAH $\rightarrow$ hcount charge $|$ charge $|$ hcount\\
symbol $\rightarrow$ aliphatic\_organic $|$ aromatic\_organic\\
isotope $\rightarrow$ DIGIT $|$ DIGIT DIGIT $|$ DIGIT DIGIT DIGIT\\
DIGIT $\rightarrow$ `1' $|$ `2' $|$ `3' $|$ `4' $|$ `5' $|$ `6' $|$ `7' $|$ `8' $|$ `9'\\
chiral $\rightarrow$ `@' $|$ `@@'\\
hcount $\rightarrow$ `H' $|$ `H' DIGIT\\
charge $\rightarrow$ `$-$' $|$ `$-$' DIGIT $|$ `+' $|$ `+' DIGIT\\
bond $\rightarrow$ `$-$' $|$ `$=$' $|$ `\#' $|$ `$/$' $|$ `\textbackslash' \\
ringbond $\rightarrow$ DIGIT $|$ bond DIGIT\\
branched\_atom $\rightarrow$ atom $|$ atom BB $|$ atom RB $|$ atom RB BB\\
RB $\rightarrow$ ringbond $|$ ringbond RB\\
BB $\rightarrow$ branch $|$ branch BB\\
branch $\rightarrow$ `(' chain `)' $|$ `(' bond chain `)' $|$ `(' AT `)' $|$ `(' chain AT `)' $|$ `(' bond chain AT `)' \\
chain $\rightarrow$ branched\_atom $|$ branched\_atom chain $|$ branched\_atom bond chain\\
}
\end{spacing}

The parse tree can be decomposed into a sequence of production rules, and can be converted into a list of 1-hot vectors, with each dimension of the vector corresponding to a rule or a terminal symbol in the polymer SMILES grammar. Thus, each polymer SMILES gets converted into matrix $\mathbf{X} \in \mathbb{R}^{T{\rm x}p}$, where p denotes the total number of production rules and terminal symbols in the polymer grammar, and T is the number of productions applied to parse the input SMILES. For consistent dimensions across different SMILES, $\mathbf{X}$ is padded with dummy numbers until $T=T_{max}$. The matrix $\mathbf{X}$ is then used as the input to the VAE encoder.

The VAE decoder attempts to reconstruct this matrix $\mathbf{X}$ or the sequence of production rules (and thus, the polymer SMILES). To ensure syntactically valid SMILES are generated, at any time during the decoding process, the decoder is only allowed to select from a subset of valid production rules. More details on the implementation of this process can be found here\cite{kusner2017grammar}.

\subsubsection{Incorporating Semantics}
The CFG implementation only guarantees production of syntactically valid SMILES. However, another crucial component for producing valid SMILES is to incorporate correct semantics or chemical rules. For example, the SMILES [*]C, [*]C($\equiv$O)C[*] and [*]c1ccccc[*] are all syntactically correct, but semantically invalid due to absence of the second polymer chain end, unrealistic triple bond with O atom and non-closure of ring bond numbered 1, respectively.

\textcolor{black}{In this work, we extend the previously proposed SD-VAE~\cite{dai2018syntax} to polymer domain. The main idea of SD-VAE is to use the concept of \emph{attribute grammar} to regulate the generation of the CFG tree by forcing the decoder to select from a smaller subset of semantically valid production rules. Formally, the attribute grammar is implemented by associating certain attribute(s) (or tags) to the non-terminal symbols of the CFG, written as $ \langle v \rangle.a$ for $v \in V$. Depending on the direction of information flow, attributes can either be \emph{inherited} (parent to child nodes) or \emph{synthesized} (child to parent nodes). However, these traditional concepts of attribute grammar can only be applied once the entire CFG tree of a molecule/polymer has been generated, making them inadequate for semantic checks. In contrast, SD-VAE proposed a generic way to introduce the concept of `lazy attribute', that allows performing the attribute grammar check alongside the generation of the CFG tree. More details on the implementation of SD-VAE for molecules is available in the original work \cite{dai2018syntax}. Here, we extend it for the case of polymers by introducing novel polymerization semantics constraint that guarantees the generated SMILES string has two chain ends with matched bond types. Furthermore, the framework extensions allow it to be applicable for both molecule and polymer discovery.}

\textcolor{black}{To realize the polymerization semantics, the following components are implemented in the generic SD-VAE framework:
\begin{itemize}
	\item \textbf{Polymer inherited attribute}: this attribute is computed in the parent-to-child direction. If during polymer generation one chain end (`[*]') has already been generated with a particular bond type, say single bond, it will constraint the next generated chain end to also have the same bond type. Further, it is used to ensure exactly two chain ends are generated.
	\item \textbf{Polymer synthesized attribute}: this attribute is computed in the child-to-parent direction. Once the first symbol `[*]' and the corresponding bond type is generated during decoding, it is used to pass this information from the respective leaf node all the way to the root node.
	\item \textbf{Polymer stochastic lazy attribute}: this attribute is exclusively introduced in the context of SD-VAE. It is used to memorize only the existence of `[*]', while delaying the calculation of the synthesized attribute, i.e., the bond type associated with `[*]'. Once the actual `[*]' is generated, its bond type will be realized, and synthesized attribute will be used to pass this information up the tree. Later, inherited attributes will be utilized for generation of the other matching `[*]'.
\end{itemize}
}

\noindent An example of attribute grammar for polymer tree generation is provided in SI. Besides these polymer semantics, we also enforced semantics common to molecular systems. Thus, overall, the following set of semantics were imposed:
\begin{itemize}
	\item Polymer chain: there should be exactly 2 `[*]' to denote the chain ends; the bond type associated with the beginning and end [*] should also match
	\item Valency constraints: limit the maximum number of bonds associated with each atom
	\item Ringbond consistency: ringbonds should come in pairs and could overlap with each other
	\item Aromatic: atom type can only be aromatic when it is part of a ring 
\end{itemize}


\subsubsection{VAE Architecture}
For the encoder, we used 3-layer one-dimensional convolution neural networks (CNNs) followed by a fully connected dense layer, output of which are used to get mean and variance of the distribution $q_{\phi}(\mathbf{z} | \mathbf{x})$ through the reparametrization trick. To decode, 3-layer gated recurrent unit (GRU) based recurrent neural network (RNN) is used followed by an affine layer activated by softmax to give probabilities $p_{\theta}(\mathbf{x} | \mathbf{z})$. The VAE architecture in inspired from the recent work on molecules\cite{gomez2018automatic,kusner2017grammar,dai2018syntax}, allowing us to compare our results with past works. The value of $T_{max}$ was set to 512, while two VAE models with latent dimensionality of 64 and 128 were trained using Adam optimization algorithm \cite{kingma2014adam}. The number of training epochs were determined by monitoring the model performance on a validation set of 20,000 SMILES taken from the hypothetical SMILES dataset. The VAE code was implemented in the PyTorch framework.

\subsection{Supervised Leaning: Gaussian Process Regression (GPR)}
After training the polymer VAE, we use GPR to learn polymer properties using their encoded latent representations. GPR uses a Bayesian framework, wherein a Gaussian process is used to obtain the functional mapping $f$($\mathbf{z}$) $\rightarrow$ $y$ based on the available training set and the Bayesian prior, incorporated using the covariance (or kernel) function. The square exponential kernel, $k(\mathbf{z},\mathbf{z'}) =  \sigma_{f}\exp \left( -\frac{1}{2\sigma^2_{l}}||\mathbf{z} -\mathbf{z'}  ||^2 \right) + \delta_{ij}\sigma^{2}_{n}$,  with three hyper-parameters $\sigma_{f}$, $\sigma_{l}$ and $\sigma_{n}$, was chosen for this work. This choice of kernel is quite standard and is known to work well for a variety of materials problems.

\subsection{Training Datasets}
Deep generative models, such as VAE, need larger datasets (in the order of 100,000 examples) 
for the training to be effective. While such large datasets are available for the case of molecules, we could only obtain information on $\sim$12,000 polymers, experimentally reported in the literature. Thus, using this initial dataset, a strategy was devised to create a larger hypothetical polymer SMILES dataset. The strategy is based on the use of retro-synthesis principles to find various polymer ``building blocks'', which are then combined together to form hypothetical polymers. Care is taken to preserve the frequency of occurrence of different building blocks in the hypothetical dataset, making it realistic and representative of the initial empirically known polymers. A total of $\sim$250,000 hypothetical polymer SMILES were created by combining various number of building blocks, ranging from 2-7 (see SI). Except 20,000 SMILES used as the validation set, all of the remaining hypothetical polymer SMILES were used for VAE training. Model parameters with lowest validation error were selected.

GPR property prediction models were trained using empirically known $T_{\rm g}$ and theoretically computed $E_{\rm g}$ datasets containing 4,997 and 963 data points, respectively. While $T_{\rm g}$ dataset was collected from experimental reports in the literature, the $E_{\rm g}$ dataset consisted of density functional theory (DFT) computed bandgap values of polymer single chains. It should be noted that measuring $E_{\rm g}$ values for polymers is a non-trivial task, and the DFT computed values can only be considered as reasonable estimates. Nonetheless, past works suggest a good correlation between the DFT computed and the empirically measured $E_{\rm g}$ values for a few common polymers \cite{chen2019electrochemical}.


\bibliographystyle{achemso}

\begin{mcitethebibliography}{54}
	\providecommand*\natexlab[1]{#1}
	\providecommand*\mciteSetBstSublistMode[1]{}
	\providecommand*\mciteSetBstMaxWidthForm[2]{}
	\providecommand*\mciteBstWouldAddEndPuncttrue
	{\def\EndOfBibitem{\unskip.}}
	\providecommand*\mciteBstWouldAddEndPunctfalse
	{\let\EndOfBibitem\relax}
	\providecommand*\mciteSetBstMidEndSepPunct[3]{}
	\providecommand*\mciteSetBstSublistLabelBeginEnd[3]{}
	\providecommand*\EndOfBibitem{}
	\mciteSetBstSublistMode{f}
	\mciteSetBstMaxWidthForm{subitem}{(\alph{mcitesubitemcount})}
	\mciteSetBstSublistLabelBeginEnd
	{\mcitemaxwidthsubitemform\space}
	{\relax}
	{\relax}
	
	\bibitem[Meredig \latin{et~al.}(2014)Meredig, Agrawal, Kirklin, Saal, Doak,
	Thompson, Zhang, Choudhary, and Wolverton]{meredig2014combinatorial}
	Meredig,~B.; Agrawal,~A.; Kirklin,~S.; Saal,~J.~E.; Doak,~J.; Thompson,~A.;
	Zhang,~K.; Choudhary,~A.; Wolverton,~C. Combinatorial screening for new
	materials in unconstrained composition space with machine learning.
	\emph{Phys. Rev. B} \textbf{2014}, \emph{89}, 094104\relax
	\mciteBstWouldAddEndPuncttrue
	\mciteSetBstMidEndSepPunct{\mcitedefaultmidpunct}
	{\mcitedefaultendpunct}{\mcitedefaultseppunct}\relax
	\EndOfBibitem
	\bibitem[Pilania \latin{et~al.}(2018)Pilania, McClellan, Stanek, and
	Uberuaga]{pilania2018physics}
	Pilania,~G.; McClellan,~K.~J.; Stanek,~C.~R.; Uberuaga,~B.~P. Physics-informed
	machine learning for inorganic scintillator discovery. \emph{J. Chem. Phys.}
	\textbf{2018}, \emph{148}, 241729\relax
	\mciteBstWouldAddEndPuncttrue
	\mciteSetBstMidEndSepPunct{\mcitedefaultmidpunct}
	{\mcitedefaultendpunct}{\mcitedefaultseppunct}\relax
	\EndOfBibitem
	\bibitem[Chen \latin{et~al.}(2019)Chen, Tran, Batra, Kim, and
	Ramprasad]{chen2019machine}
	Chen,~L.; Tran,~H.; Batra,~R.; Kim,~C.; Ramprasad,~R. Machine learning models
	for the lattice thermal conductivity prediction of inorganic materials.
	\emph{Comput. Mater. Sci.} \textbf{2019}, \emph{170}, 109155\relax
	\mciteBstWouldAddEndPuncttrue
	\mciteSetBstMidEndSepPunct{\mcitedefaultmidpunct}
	{\mcitedefaultendpunct}{\mcitedefaultseppunct}\relax
	\EndOfBibitem
	\bibitem[Tshitoyan \latin{et~al.}(2019)Tshitoyan, Dagdelen, Weston, Dunn, Rong,
	Kononova, Persson, Ceder, and Jain]{tshitoyan2019unsupervised}
	Tshitoyan,~V.; Dagdelen,~J.; Weston,~L.; Dunn,~A.; Rong,~Z.; Kononova,~O.;
	Persson,~K.~A.; Ceder,~G.; Jain,~A. Unsupervised word embeddings capture
	latent knowledge from materials science literature. \emph{Nature}
	\textbf{2019}, \emph{571}, 95\relax
	\mciteBstWouldAddEndPuncttrue
	\mciteSetBstMidEndSepPunct{\mcitedefaultmidpunct}
	{\mcitedefaultendpunct}{\mcitedefaultseppunct}\relax
	\EndOfBibitem
	\bibitem[Sun \latin{et~al.}(2019)Sun, Ouyang, Zhang, and Zhang]{sun2019data}
	Sun,~S.; Ouyang,~R.; Zhang,~B.; Zhang,~T.-Y. Data-driven discovery of formulas
	by symbolic regression. \emph{MRS Bull.} \textbf{2019}, \emph{44},
	559--564\relax
	\mciteBstWouldAddEndPuncttrue
	\mciteSetBstMidEndSepPunct{\mcitedefaultmidpunct}
	{\mcitedefaultendpunct}{\mcitedefaultseppunct}\relax
	\EndOfBibitem
	\bibitem[Vasudevan \latin{et~al.}(2019)Vasudevan, Choudhary, Mehta, Smith,
	Kusne, Tavazza, Vlcek, Ziatdinov, Kalinin, and
	Hattrick-Simpers]{vasudevan2019materials}
	Vasudevan,~R.~K.; Choudhary,~K.; Mehta,~A.; Smith,~R.; Kusne,~G.; Tavazza,~F.;
	Vlcek,~L.; Ziatdinov,~M.; Kalinin,~S.~V.; Hattrick-Simpers,~J. Materials
	science in the artificial intelligence age: high-throughput library
	generation, machine learning, and a pathway from correlations to the
	underpinning physics. \emph{MRS Commun.} \textbf{2019}, 1--18\relax
	\mciteBstWouldAddEndPuncttrue
	\mciteSetBstMidEndSepPunct{\mcitedefaultmidpunct}
	{\mcitedefaultendpunct}{\mcitedefaultseppunct}\relax
	\EndOfBibitem
	\bibitem[Kim \latin{et~al.}(2017)Kim, Huang, Saunders, McCallum, Ceder, and
	Olivetti]{kim2017materials}
	Kim,~E.; Huang,~K.; Saunders,~A.; McCallum,~A.; Ceder,~G.; Olivetti,~E.
	Materials synthesis insights from scientific literature via text extraction
	and machine learning. \emph{Chem. Mater.} \textbf{2017}, \emph{29},
	9436--9444\relax
	\mciteBstWouldAddEndPuncttrue
	\mciteSetBstMidEndSepPunct{\mcitedefaultmidpunct}
	{\mcitedefaultendpunct}{\mcitedefaultseppunct}\relax
	\EndOfBibitem
	\bibitem[King \latin{et~al.}(2009)King, Rowland, Oliver, Young, Aubrey, Byrne,
	Liakata, Markham, Pir, Soldatova, Sparkes, Whelan, and
	Clare]{king2009automation}
	King,~R.~D.; Rowland,~J.; Oliver,~S.~G.; Young,~M.; Aubrey,~W.; Byrne,~E.;
	Liakata,~M.; Markham,~M.; Pir,~P.; Soldatova,~L.~N.; Sparkes,~A.;
	Whelan,~K.~E.; Clare,~A. The automation of science. \emph{Science}
	\textbf{2009}, \emph{324}, 85--89\relax
	\mciteBstWouldAddEndPuncttrue
	\mciteSetBstMidEndSepPunct{\mcitedefaultmidpunct}
	{\mcitedefaultendpunct}{\mcitedefaultseppunct}\relax
	\EndOfBibitem
	\bibitem[Kim \latin{et~al.}(2018)Kim, Chandrasekaran, Huan, Das, and
	Ramprasad]{kim2018polymer}
	Kim,~C.; Chandrasekaran,~A.; Huan,~T.~D.; Das,~D.; Ramprasad,~R. Polymer
	genome: a data-powered polymer informatics platform for property predictions.
	\emph{J. Phys. Chem. C} \textbf{2018}, \emph{122}, 17575--17585\relax
	\mciteBstWouldAddEndPuncttrue
	\mciteSetBstMidEndSepPunct{\mcitedefaultmidpunct}
	{\mcitedefaultendpunct}{\mcitedefaultseppunct}\relax
	\EndOfBibitem
	\bibitem[Xue \latin{et~al.}(2016)Xue, Balachandran, Hogden, Theiler, Xue, and
	Lookman]{xue2016accelerated}
	Xue,~D.; Balachandran,~P.~V.; Hogden,~J.; Theiler,~J.; Xue,~D.; Lookman,~T.
	Accelerated search for materials with targeted properties by adaptive design.
	\emph{Nat. Commun.} \textbf{2016}, \emph{7}, 11241\relax
	\mciteBstWouldAddEndPuncttrue
	\mciteSetBstMidEndSepPunct{\mcitedefaultmidpunct}
	{\mcitedefaultendpunct}{\mcitedefaultseppunct}\relax
	\EndOfBibitem
	\bibitem[Mansouri~Tehrani \latin{et~al.}(2018)Mansouri~Tehrani, Oliynyk, Parry,
	Rizvi, Couper, Lin, Miyagi, Sparks, and Brgoch]{mansouri2018machine}
	Mansouri~Tehrani,~A.; Oliynyk,~A.~O.; Parry,~M.; Rizvi,~Z.; Couper,~S.;
	Lin,~F.; Miyagi,~L.; Sparks,~T.~D.; Brgoch,~J. Machine learning directed
	search for ultraincompressible, superhard materials. \emph{J. Am. Chem. Soc.}
	\textbf{2018}, \emph{140}, 9844--9853\relax
	\mciteBstWouldAddEndPuncttrue
	\mciteSetBstMidEndSepPunct{\mcitedefaultmidpunct}
	{\mcitedefaultendpunct}{\mcitedefaultseppunct}\relax
	\EndOfBibitem
	\bibitem[Ren \latin{et~al.}(2018)Ren, Ward, Williams, Laws, Wolverton,
	Hattrick-Simpers, and Mehta]{ren2018accelerated}
	Ren,~F.; Ward,~L.; Williams,~T.; Laws,~K.~J.; Wolverton,~C.;
	Hattrick-Simpers,~J.; Mehta,~A. Accelerated discovery of metallic glasses
	through iteration of machine learning and high-throughput experiments.
	\emph{Sci. Adv.} \textbf{2018}, \emph{4}, eaaq1566\relax
	\mciteBstWouldAddEndPuncttrue
	\mciteSetBstMidEndSepPunct{\mcitedefaultmidpunct}
	{\mcitedefaultendpunct}{\mcitedefaultseppunct}\relax
	\EndOfBibitem
	\bibitem[Stanev \latin{et~al.}(2018)Stanev, Oses, Kusne, Rodriguez, Paglione,
	Curtarolo, and Takeuchi]{stanev2018machine}
	Stanev,~V.; Oses,~C.; Kusne,~A.~G.; Rodriguez,~E.; Paglione,~J.; Curtarolo,~S.;
	Takeuchi,~I. Machine learning modeling of superconducting critical
	temperature. \emph{npj Comput. Mater.} \textbf{2018}, \emph{4}, 29\relax
	\mciteBstWouldAddEndPuncttrue
	\mciteSetBstMidEndSepPunct{\mcitedefaultmidpunct}
	{\mcitedefaultendpunct}{\mcitedefaultseppunct}\relax
	\EndOfBibitem
	\bibitem[Lopez \latin{et~al.}(2017)Lopez, Sanchez-Lengeling, de~Goes~Soares,
	and Aspuru-Guzik]{lopez2017design}
	Lopez,~S.~A.; Sanchez-Lengeling,~B.; de~Goes~Soares,~J.; Aspuru-Guzik,~A.
	Design principles and top non-fullerene acceptor candidates for organic
	photovoltaics. \emph{Joule} \textbf{2017}, \emph{1}, 857--870\relax
	\mciteBstWouldAddEndPuncttrue
	\mciteSetBstMidEndSepPunct{\mcitedefaultmidpunct}
	{\mcitedefaultendpunct}{\mcitedefaultseppunct}\relax
	\EndOfBibitem
	\bibitem[Kanal \latin{et~al.}(2013)Kanal, Owens, Bechtel, and
	Hutchison]{kanal2013efficient}
	Kanal,~I.~Y.; Owens,~S.~G.; Bechtel,~J.~S.; Hutchison,~G.~R. Efficient
	computational screening of organic polymer photovoltaics. \emph{J. Phys.
		Chem. Lett.} \textbf{2013}, \emph{4}, 1613--1623\relax
	\mciteBstWouldAddEndPuncttrue
	\mciteSetBstMidEndSepPunct{\mcitedefaultmidpunct}
	{\mcitedefaultendpunct}{\mcitedefaultseppunct}\relax
	\EndOfBibitem
	\bibitem[Cheng \latin{et~al.}(2015)Cheng, Assary, Qu, Jain, Ong, Rajput,
	Persson, and Curtiss]{cheng2015accelerating}
	Cheng,~L.; Assary,~R.~S.; Qu,~X.; Jain,~A.; Ong,~S.~P.; Rajput,~N.~N.;
	Persson,~K.; Curtiss,~L.~A. Accelerating electrolyte discovery for energy
	storage with high-throughput screening. \emph{J. Phys. Chem. Lett.}
	\textbf{2015}, \emph{6}, 283--291\relax
	\mciteBstWouldAddEndPuncttrue
	\mciteSetBstMidEndSepPunct{\mcitedefaultmidpunct}
	{\mcitedefaultendpunct}{\mcitedefaultseppunct}\relax
	\EndOfBibitem
	\bibitem[Chandrasekaran \latin{et~al.}(2019)Chandrasekaran, Kamal, Batra, Kim,
	Chen, and Ramprasad]{chandrasekaran2019solving}
	Chandrasekaran,~A.; Kamal,~D.; Batra,~R.; Kim,~C.; Chen,~L.; Ramprasad,~R.
	Solving the electronic structure problem with machine learning. \emph{npj
		Comput. Mater.} \textbf{2019}, \emph{5}, 22\relax
	\mciteBstWouldAddEndPuncttrue
	\mciteSetBstMidEndSepPunct{\mcitedefaultmidpunct}
	{\mcitedefaultendpunct}{\mcitedefaultseppunct}\relax
	\EndOfBibitem
	\bibitem[Huan \latin{et~al.}(2017)Huan, Batra, Chapman, Krishnan, Chen, and
	Ramprasad]{huan2017universal}
	Huan,~T.~D.; Batra,~R.; Chapman,~J.; Krishnan,~S.; Chen,~L.; Ramprasad,~R. A
	universal strategy for the creation of machine learning-based atomistic force
	fields. \emph{npj Comput. Mater.} \textbf{2017}, \emph{3}, 37\relax
	\mciteBstWouldAddEndPuncttrue
	\mciteSetBstMidEndSepPunct{\mcitedefaultmidpunct}
	{\mcitedefaultendpunct}{\mcitedefaultseppunct}\relax
	\EndOfBibitem
	\bibitem[Pun \latin{et~al.}(2019)Pun, Batra, Ramprasad, and
	Mishin]{pun2019physically}
	Pun,~G.~P.; Batra,~R.; Ramprasad,~R.; Mishin,~Y. Physically informed artificial
	neural networks for atomistic modeling of materials. \emph{Nat. Commun.}
	\textbf{2019}, \emph{10}\relax
	\mciteBstWouldAddEndPuncttrue
	\mciteSetBstMidEndSepPunct{\mcitedefaultmidpunct}
	{\mcitedefaultendpunct}{\mcitedefaultseppunct}\relax
	\EndOfBibitem
	\bibitem[Huan \latin{et~al.}(2019)Huan, Batra, Chapman, Kim, Chandrasekaran,
	and Ramprasad]{huan2019iterative}
	Huan,~T.~D.; Batra,~R.; Chapman,~J.; Kim,~C.; Chandrasekaran,~A.; Ramprasad,~R.
	Iterative-learning strategy for the development of application-specific
	atomistic force fields. \emph{J. Phys. Chem. C} \textbf{2019}, \emph{123},
	20715--20722\relax
	\mciteBstWouldAddEndPuncttrue
	\mciteSetBstMidEndSepPunct{\mcitedefaultmidpunct}
	{\mcitedefaultendpunct}{\mcitedefaultseppunct}\relax
	\EndOfBibitem
	\bibitem[{Shahriari} \latin{et~al.}(2016){Shahriari}, {Swersky}, {Wang},
	{Adams}, and {de Freitas}]{bayesianlearning}
	{Shahriari},~B.; {Swersky},~K.; {Wang},~Z.; {Adams},~R.~P.; {de Freitas},~N.
	Taking the human out of the loop: A review of bayesian optimization.
	\emph{Proceedings of the IEEE} \textbf{2016}, \emph{104}, 148--175\relax
	\mciteBstWouldAddEndPuncttrue
	\mciteSetBstMidEndSepPunct{\mcitedefaultmidpunct}
	{\mcitedefaultendpunct}{\mcitedefaultseppunct}\relax
	\EndOfBibitem
	\bibitem[Kim \latin{et~al.}(2019)Kim, Chandrasekaran, Jha, and
	Ramprasad]{kim2019active}
	Kim,~C.; Chandrasekaran,~A.; Jha,~A.; Ramprasad,~R. Active-learning and
	materials design: the example of high glass transition temperature polymers.
	\emph{MRS Commun.} \textbf{2019}, 1--7\relax
	\mciteBstWouldAddEndPuncttrue
	\mciteSetBstMidEndSepPunct{\mcitedefaultmidpunct}
	{\mcitedefaultendpunct}{\mcitedefaultseppunct}\relax
	\EndOfBibitem
	\bibitem[Walsh(2015)]{walsh2015inorganic}
	Walsh,~A. Inorganic materials: The quest for new functionality. \emph{Nat.
		Chem.} \textbf{2015}, \emph{7}, 274\relax
	\mciteBstWouldAddEndPuncttrue
	\mciteSetBstMidEndSepPunct{\mcitedefaultmidpunct}
	{\mcitedefaultendpunct}{\mcitedefaultseppunct}\relax
	\EndOfBibitem
	\bibitem[Kim \latin{et~al.}(2021)Kim, Batra, Chen, Tran, and
	Ramprasad]{kim2021polymer}
	Kim,~C.; Batra,~R.; Chen,~L.; Tran,~H.; Ramprasad,~R. Polymer design using
	genetic algorithm and machine learning. \emph{Comput. Mater. Sci.}
	\textbf{2021}, \emph{186}, 110067\relax
	\mciteBstWouldAddEndPuncttrue
	\mciteSetBstMidEndSepPunct{\mcitedefaultmidpunct}
	{\mcitedefaultendpunct}{\mcitedefaultseppunct}\relax
	\EndOfBibitem
	\bibitem[G{\'o}mez-Bombarelli \latin{et~al.}(2018)G{\'o}mez-Bombarelli, Wei,
	Duvenaud, Hern{\'a}ndez-Lobato, S{\'a}nchez-Lengeling, Sheberla,
	Aguilera-Iparraguirre, Hirzel, Adams, and Aspuru-Guzik]{gomez2018automatic}
	G{\'o}mez-Bombarelli,~R.; Wei,~J.~N.; Duvenaud,~D.;
	Hern{\'a}ndez-Lobato,~J.~M.; S{\'a}nchez-Lengeling,~B.; Sheberla,~D.;
	Aguilera-Iparraguirre,~J.; Hirzel,~T.~D.; Adams,~R.~P.; Aspuru-Guzik,~A.
	Automatic chemical design using a data-driven continuous representation of
	molecules. \emph{ACS Cent. Sci.} \textbf{2018}, \emph{4}, 268--276\relax
	\mciteBstWouldAddEndPuncttrue
	\mciteSetBstMidEndSepPunct{\mcitedefaultmidpunct}
	{\mcitedefaultendpunct}{\mcitedefaultseppunct}\relax
	\EndOfBibitem
	\bibitem[Kusner \latin{et~al.}(2017)Kusner, Paige, and
	Hern{\'a}ndez-Lobato]{kusner2017grammar}
	Kusner,~M.~J.; Paige,~B.; Hern{\'a}ndez-Lobato,~J.~M. Grammar variational
	autoencoder. Proceedings of the 34th International Conference on Machine
	Learning-Volume 70. 2017; pp 1945--1954\relax
	\mciteBstWouldAddEndPuncttrue
	\mciteSetBstMidEndSepPunct{\mcitedefaultmidpunct}
	{\mcitedefaultendpunct}{\mcitedefaultseppunct}\relax
	\EndOfBibitem
	\bibitem[Dai \latin{et~al.}(2018)Dai, Tian, Dai, Skiena, and
	Song]{dai2018syntax}
	Dai,~H.; Tian,~Y.; Dai,~B.; Skiena,~S.; Song,~L. Syntax-directed variational
	autoencoder for structured data. \emph{arXiv:1802.08786} \textbf{2018},
	\relax
	\mciteBstWouldAddEndPunctfalse
	\mciteSetBstMidEndSepPunct{\mcitedefaultmidpunct}
	{}{\mcitedefaultseppunct}\relax
	\EndOfBibitem
	\bibitem[Jin \latin{et~al.}(2018)Jin, Barzilay, and Jaakkola]{jin2018junction}
	Jin,~W.; Barzilay,~R.; Jaakkola,~T. Junction tree variational autoencoder for
	molecular graph generation. \emph{arXiv:1802.04364} \textbf{2018}, \relax
	\mciteBstWouldAddEndPunctfalse
	\mciteSetBstMidEndSepPunct{\mcitedefaultmidpunct}
	{}{\mcitedefaultseppunct}\relax
	\EndOfBibitem
	\bibitem[Liu \latin{et~al.}(2018)Liu, Allamanis, Brockschmidt, and
	Gaunt]{liu2018constrained}
	Liu,~Q.; Allamanis,~M.; Brockschmidt,~M.; Gaunt,~A. Constrained graph
	variational autoencoders for molecule design. Adv. Neural Inf. Process. Syst.
	2018; pp 7795--7804\relax
	\mciteBstWouldAddEndPuncttrue
	\mciteSetBstMidEndSepPunct{\mcitedefaultmidpunct}
	{\mcitedefaultendpunct}{\mcitedefaultseppunct}\relax
	\EndOfBibitem
	\bibitem[Guimaraes \latin{et~al.}(2017)Guimaraes, Sanchez-Lengeling, Outeiral,
	Farias, and Aspuru-Guzik]{guimaraes2017objective}
	Guimaraes,~G.~L.; Sanchez-Lengeling,~B.; Outeiral,~C.; Farias,~P. L.~C.;
	Aspuru-Guzik,~A. Objective-reinforced generative adversarial networks (ORGAN)
	for sequence generation models. \emph{arXiv:1705.10843} \textbf{2017}, \relax
	\mciteBstWouldAddEndPunctfalse
	\mciteSetBstMidEndSepPunct{\mcitedefaultmidpunct}
	{}{\mcitedefaultseppunct}\relax
	\EndOfBibitem
	\bibitem[You \latin{et~al.}(2018)You, Liu, Ying, Pande, and
	Leskovec]{you2018graph}
	You,~J.; Liu,~B.; Ying,~Z.; Pande,~V.; Leskovec,~J. Graph convolutional policy
	network for goal-directed molecular graph generation. Adv. Neural Inf.
	Process. Syst. 2018; pp 6410--6421\relax
	\mciteBstWouldAddEndPuncttrue
	\mciteSetBstMidEndSepPunct{\mcitedefaultmidpunct}
	{\mcitedefaultendpunct}{\mcitedefaultseppunct}\relax
	\EndOfBibitem
	\bibitem[De~Cao and Kipf(2018)De~Cao, and Kipf]{de2018molgan}
	De~Cao,~N.; Kipf,~T. MolGAN: An implicit generative model for small molecular
	graphs. \emph{arXiv:1805.11973} \textbf{2018}, \relax
	\mciteBstWouldAddEndPunctfalse
	\mciteSetBstMidEndSepPunct{\mcitedefaultmidpunct}
	{}{\mcitedefaultseppunct}\relax
	\EndOfBibitem
	\bibitem[Kotsias \latin{et~al.}(2020)Kotsias, Ar{\'u}s-Pous, Chen, Engkvist,
	Tyrchan, and Bjerrum]{kotsias2020direct}
	Kotsias,~P.-C.; Ar{\'u}s-Pous,~J.; Chen,~H.; Engkvist,~O.; Tyrchan,~C.;
	Bjerrum,~E.~J. Direct steering of de novo molecular generation with
	descriptor conditional recurrent neural networks. \emph{Nat. Mach. Intell.}
	\textbf{2020}, \emph{2}, 254--265\relax
	\mciteBstWouldAddEndPuncttrue
	\mciteSetBstMidEndSepPunct{\mcitedefaultmidpunct}
	{\mcitedefaultendpunct}{\mcitedefaultseppunct}\relax
	\EndOfBibitem
	\bibitem[Krenn \latin{et~al.}(2020)Krenn, Hase, Nigam, Friederich, and
	Aspuru-Guzik]{krenn2020self}
	Krenn,~M.; Hase,~F.; Nigam,~A.; Friederich,~P.; Aspuru-Guzik,~A.
	Self-Referencing Embedded Strings {(SELFIES)}: A 100\% robust molecular
	string representation. \emph{Mach. Learn. Sci. Tech.} \textbf{2020}, \relax
	\mciteBstWouldAddEndPunctfalse
	\mciteSetBstMidEndSepPunct{\mcitedefaultmidpunct}
	{}{\mcitedefaultseppunct}\relax
	\EndOfBibitem
	\bibitem[J{\o}rgensen \latin{et~al.}(2018)J{\o}rgensen, Mesta, Shil,
	Garc{\'\i}a~Lastra, Jacobsen, Thygesen, and Schmidt]{jorgensen2018machine}
	J{\o}rgensen,~P.~B.; Mesta,~M.; Shil,~S.; Garc{\'\i}a~Lastra,~J.~M.;
	Jacobsen,~K.~W.; Thygesen,~K.~S.; Schmidt,~M.~N. Machine learning-based
	screening of complex molecules for polymer solar cells. \emph{J. Chem. Phys.}
	\textbf{2018}, \emph{148}, 241735\relax
	\mciteBstWouldAddEndPuncttrue
	\mciteSetBstMidEndSepPunct{\mcitedefaultmidpunct}
	{\mcitedefaultendpunct}{\mcitedefaultseppunct}\relax
	\EndOfBibitem
	\bibitem[Kim \latin{et~al.}(2020)Kim, Lee, and Kim]{kim2020inverse}
	Kim,~B.; Lee,~S.; Kim,~J. Inverse design of porous materials using artificial
	neural networks. \emph{Sci. Adv.} \textbf{2020}, \emph{6}, eaax9324\relax
	\mciteBstWouldAddEndPuncttrue
	\mciteSetBstMidEndSepPunct{\mcitedefaultmidpunct}
	{\mcitedefaultendpunct}{\mcitedefaultseppunct}\relax
	\EndOfBibitem
	\bibitem[Allamanis \latin{et~al.}(2017)Allamanis, Chanthirasegaran, Kohli, and
	Sutton]{allamanis2017learning}
	Allamanis,~M.; Chanthirasegaran,~P.; Kohli,~P.; Sutton,~C. Learning continuous
	semantic representations of symbolic expressions. Proceedings of the 34th
	International Conference on Machine Learning-Volume 70. 2017; pp 80--88\relax
	\mciteBstWouldAddEndPuncttrue
	\mciteSetBstMidEndSepPunct{\mcitedefaultmidpunct}
	{\mcitedefaultendpunct}{\mcitedefaultseppunct}\relax
	\EndOfBibitem
	\bibitem[Kernighan \latin{et~al.}(1988)Kernighan, Ritchie, and
	Ejeklint]{cbook1988}
	Kernighan,~B.~W.; Ritchie,~D.~M.; Ejeklint,~P. \emph{The C programming
		language, volume 2}; Prentice-Hall Englewood Cliffs, 1988\relax
	\mciteBstWouldAddEndPuncttrue
	\mciteSetBstMidEndSepPunct{\mcitedefaultmidpunct}
	{\mcitedefaultendpunct}{\mcitedefaultseppunct}\relax
	\EndOfBibitem
	\bibitem[St.~John \latin{et~al.}(2019)St.~John, Phillips, Kemper, Wilson, Guan,
	Crowley, Nimlos, and Larsen]{st2019message}
	St.~John,~P.~C.; Phillips,~C.; Kemper,~T.~W.; Wilson,~A.~N.; Guan,~Y.;
	Crowley,~M.~F.; Nimlos,~M.~R.; Larsen,~R.~E. Message-passing neural networks
	for high-throughput polymer screening. \emph{J. Chem. Phys.} \textbf{2019},
	\emph{150}, 234111\relax
	\mciteBstWouldAddEndPuncttrue
	\mciteSetBstMidEndSepPunct{\mcitedefaultmidpunct}
	{\mcitedefaultendpunct}{\mcitedefaultseppunct}\relax
	\EndOfBibitem
	\bibitem[Jin \latin{et~al.}(2020)Jin, Barzilay, and
	Jaakkola]{jin2020hierarchical}
	Jin,~W.; Barzilay,~R.; Jaakkola,~T. Hierarchical Generation of Molecular Graphs
	using Structural Motifs. \emph{arXiv preprint arXiv:2002.03230}
	\textbf{2020}, \relax
	\mciteBstWouldAddEndPunctfalse
	\mciteSetBstMidEndSepPunct{\mcitedefaultmidpunct}
	{}{\mcitedefaultseppunct}\relax
	\EndOfBibitem
	\bibitem[Ho and Greenbaum(2018)Ho, and Greenbaum]{ho2018polymer}
	Ho,~J.~S.; Greenbaum,~S.~G. Polymer capacitor dielectrics for high temperature
	applications. \emph{ACS Appl. Mater. Interfaces} \textbf{2018}, \emph{10},
	29189--29218\relax
	\mciteBstWouldAddEndPuncttrue
	\mciteSetBstMidEndSepPunct{\mcitedefaultmidpunct}
	{\mcitedefaultendpunct}{\mcitedefaultseppunct}\relax
	\EndOfBibitem
	\bibitem[Tan \latin{et~al.}(2014)Tan, Zhang, Chen, and Irwin]{tan2014high}
	Tan,~D.; Zhang,~L.; Chen,~Q.; Irwin,~P. High-temperature capacitor polymer
	films. \emph{J. Electron. Mater.} \textbf{2014}, \emph{43}, 4569--4575\relax
	\mciteBstWouldAddEndPuncttrue
	\mciteSetBstMidEndSepPunct{\mcitedefaultmidpunct}
	{\mcitedefaultendpunct}{\mcitedefaultseppunct}\relax
	\EndOfBibitem
	\bibitem[Zhou \latin{et~al.}(2018)Zhou, Li, Dang, Yang, Shao, Li, Hu, Zeng, He,
	and Wang]{zhou2018scalable}
	Zhou,~Y.; Li,~Q.; Dang,~B.; Yang,~Y.; Shao,~T.; Li,~H.; Hu,~J.; Zeng,~R.;
	He,~J.; Wang,~Q. A scalable, high-throughput, and environmentally benign
	approach to polymer dielectrics exhibiting significantly improved capacitive
	performance at high temperatures. \emph{Adv. Mater.} \textbf{2018},
	\emph{30}, 1805672\relax
	\mciteBstWouldAddEndPuncttrue
	\mciteSetBstMidEndSepPunct{\mcitedefaultmidpunct}
	{\mcitedefaultendpunct}{\mcitedefaultseppunct}\relax
	\EndOfBibitem
	\bibitem[Wu \latin{et~al.}(2020)Wu, Deshmukh, Li, Chen, Alamri, Wang,
	Ramprasad, Sotzing, and Cao]{wu2020flexible}
	Wu,~C.; Deshmukh,~A.~A.; Li,~Z.; Chen,~L.; Alamri,~A.; Wang,~Y.; Ramprasad,~R.;
	Sotzing,~G.~A.; Cao,~Y. Flexible temperature-invariant polymer dielectrics
	with large bandgap. \emph{Adv. Mater.} \textbf{2020}, \emph{32},
	2000499\relax
	\mciteBstWouldAddEndPuncttrue
	\mciteSetBstMidEndSepPunct{\mcitedefaultmidpunct}
	{\mcitedefaultendpunct}{\mcitedefaultseppunct}\relax
	\EndOfBibitem
	\bibitem[Mannodi-Kanakkithodi \latin{et~al.}(2018)Mannodi-Kanakkithodi,
	Chandrasekaran, Kim, Huan, Pilania, Botu, and Ramprasad]{mannodi2018scoping}
	Mannodi-Kanakkithodi,~A.; Chandrasekaran,~A.; Kim,~C.; Huan,~T.~D.;
	Pilania,~G.; Botu,~V.; Ramprasad,~R. Scoping the polymer genome: A roadmap
	for rational polymer dielectrics design and beyond. \emph{Mater. Today}
	\textbf{2018}, \emph{21}, 785--796\relax
	\mciteBstWouldAddEndPuncttrue
	\mciteSetBstMidEndSepPunct{\mcitedefaultmidpunct}
	{\mcitedefaultendpunct}{\mcitedefaultseppunct}\relax
	\EndOfBibitem
	\bibitem[Kim \latin{et~al.}(2016)Kim, Pilania, and Ramprasad]{kim2016organized}
	Kim,~C.; Pilania,~G.; Ramprasad,~R. From organized high-throughput data to
	phenomenological theory using machine learning: the example of dielectric
	breakdown. \emph{Chem. Mater.} \textbf{2016}, \emph{28}, 1304--1311\relax
	\mciteBstWouldAddEndPuncttrue
	\mciteSetBstMidEndSepPunct{\mcitedefaultmidpunct}
	{\mcitedefaultendpunct}{\mcitedefaultseppunct}\relax
	\EndOfBibitem
	\bibitem[Kamal \latin{et~al.}(2020)Kamal, Wang, Tran, Chen, Li, Wu, Nasreen,
	Cao, and Ramprasad]{kamal2020computable}
	Kamal,~D.; Wang,~Y.; Tran,~H.~D.; Chen,~L.; Li,~Z.; Wu,~C.; Nasreen,~S.;
	Cao,~Y.; Ramprasad,~R. Computable bulk and interfacial electronic structure
	features as proxies for dielectric breakdown of polymers. \emph{ACS Appl.
		Mater. Interfaces} \textbf{2020}, PMID: 32705867\relax
	\mciteBstWouldAddEndPuncttrue
	\mciteSetBstMidEndSepPunct{\mcitedefaultmidpunct}
	{\mcitedefaultendpunct}{\mcitedefaultseppunct}\relax
	\EndOfBibitem
	\bibitem[Lin \latin{et~al.}(2019)Lin, Coley, Mochigase, Beech, Wang, Wang,
	Woods, Craig, Johnson, Kalow, \latin{et~al.} others]{lin2019bigsmiles}
	Lin,~T.-S.; Coley,~C.~W.; Mochigase,~H.; Beech,~H.~K.; Wang,~W.; Wang,~Z.;
	Woods,~E.; Craig,~S.~L.; Johnson,~J.~A.; Kalow,~J.~A., \latin{et~al.}
	BigSMILES: a structurally-based line notation for describing macromolecules.
	\emph{ACS Cent. Sci.} \textbf{2019}, \emph{5}, 1523--1531\relax
	\mciteBstWouldAddEndPuncttrue
	\mciteSetBstMidEndSepPunct{\mcitedefaultmidpunct}
	{\mcitedefaultendpunct}{\mcitedefaultseppunct}\relax
	\EndOfBibitem
	\bibitem[Zhou \latin{et~al.}(2019)Zhou, Kearnes, Li, Zare, and
	Riley]{zhou2019optimization}
	Zhou,~Z.; Kearnes,~S.; Li,~L.; Zare,~R.~N.; Riley,~P. Optimization of molecules
	via deep reinforcement learning. \emph{Sci. Rep.} \textbf{2019}, \emph{9},
	1--10\relax
	\mciteBstWouldAddEndPuncttrue
	\mciteSetBstMidEndSepPunct{\mcitedefaultmidpunct}
	{\mcitedefaultendpunct}{\mcitedefaultseppunct}\relax
	\EndOfBibitem
	\bibitem[Kingma and Welling(2014)Kingma, and Welling]{kingma2014auto}
	Kingma,~D.~P.; Welling,~M. Auto-encoding variational bayes. In Proceedings of
	the International Conference on Learning Representations (ICLR). 2014\relax
	\mciteBstWouldAddEndPuncttrue
	\mciteSetBstMidEndSepPunct{\mcitedefaultmidpunct}
	{\mcitedefaultendpunct}{\mcitedefaultseppunct}\relax
	\EndOfBibitem
	\bibitem[Hopcroft \latin{et~al.}(2006)Hopcroft, Motwani, and
	Ullman]{grammarbook2006}
	Hopcroft,~J.~E.; Motwani,~R.; Ullman,~J.~D. \emph{Introduction to Automata
		theory, languages and computation}; Addison-Wesley, 2006\relax
	\mciteBstWouldAddEndPuncttrue
	\mciteSetBstMidEndSepPunct{\mcitedefaultmidpunct}
	{\mcitedefaultendpunct}{\mcitedefaultseppunct}\relax
	\EndOfBibitem
	\bibitem[Kingma and Ba(2015)Kingma, and Ba]{kingma2014adam}
	Kingma,~D.~P.; Ba,~J. Adam: A method for stochastic optimization. International
	Conference on Learning Representations. 2015\relax
	\mciteBstWouldAddEndPuncttrue
	\mciteSetBstMidEndSepPunct{\mcitedefaultmidpunct}
	{\mcitedefaultendpunct}{\mcitedefaultseppunct}\relax
	\EndOfBibitem
	\bibitem[Chen \latin{et~al.}(2019)Chen, Venkatram, Kim, Batra, Chandrasekaran,
	and Ramprasad]{chen2019electrochemical}
	Chen,~L.; Venkatram,~S.; Kim,~C.; Batra,~R.; Chandrasekaran,~A.; Ramprasad,~R.
	Electrochemical stability window of polymeric electrolytes. \emph{Chem.
		Mater.} \textbf{2019}, \emph{31}, 4598--4604\relax
	\mciteBstWouldAddEndPuncttrue
	\mciteSetBstMidEndSepPunct{\mcitedefaultmidpunct}
	{\mcitedefaultendpunct}{\mcitedefaultseppunct}\relax
	\EndOfBibitem
\end{mcitethebibliography}


\providecommand{\latin}[1]{#1}
\makeatletter
\providecommand{\doi}
{\begingroup\let\do\@makeother\dospecials
	\catcode`\{=1 \catcode`\}=2 \doi@aux}
\providecommand{\doi@aux}[1]{\endgroup\texttt{#1}}
\makeatother
\providecommand*\mcitethebibliography{\thebibliography}
\csname @ifundefined\endcsname{endmcitethebibliography}
{\let\endmcitethebibliography\endthebibliography}{}





\section{Additional information}
\section{Acknowledgements}
This work is supported by the Office of Naval Research through N0014-17-1-2656, a Multi-University Research Initiative (MURI) grant. R.B. acknowledges the use of the Center for Nanoscale Materials, an Office of Science user facility, and support by the U.S. Department of Energy, Office of Science, Office of Basic Energy Sciences, under Contract No. DE-AC02-06CH11357.

\subsection{Supplementary Information}
List of the new polymers (SMILES and VAE property predictions) designed in this work, and additional discussion on the VAE-128 latent space, polymer bandgap validation, hypothetical SMILES data set, chemical fragment analysis and polymer attribute grammar.

\subsection{Competing interests} 
The authors declare no competing interests.

\subsection{Keywords}
variational autoencoders, polymer design, machine learning, polymer dielectrics

\section{TOC}
\begin{figure}
	\begin{center}
		\includegraphics[width=8.47cm]{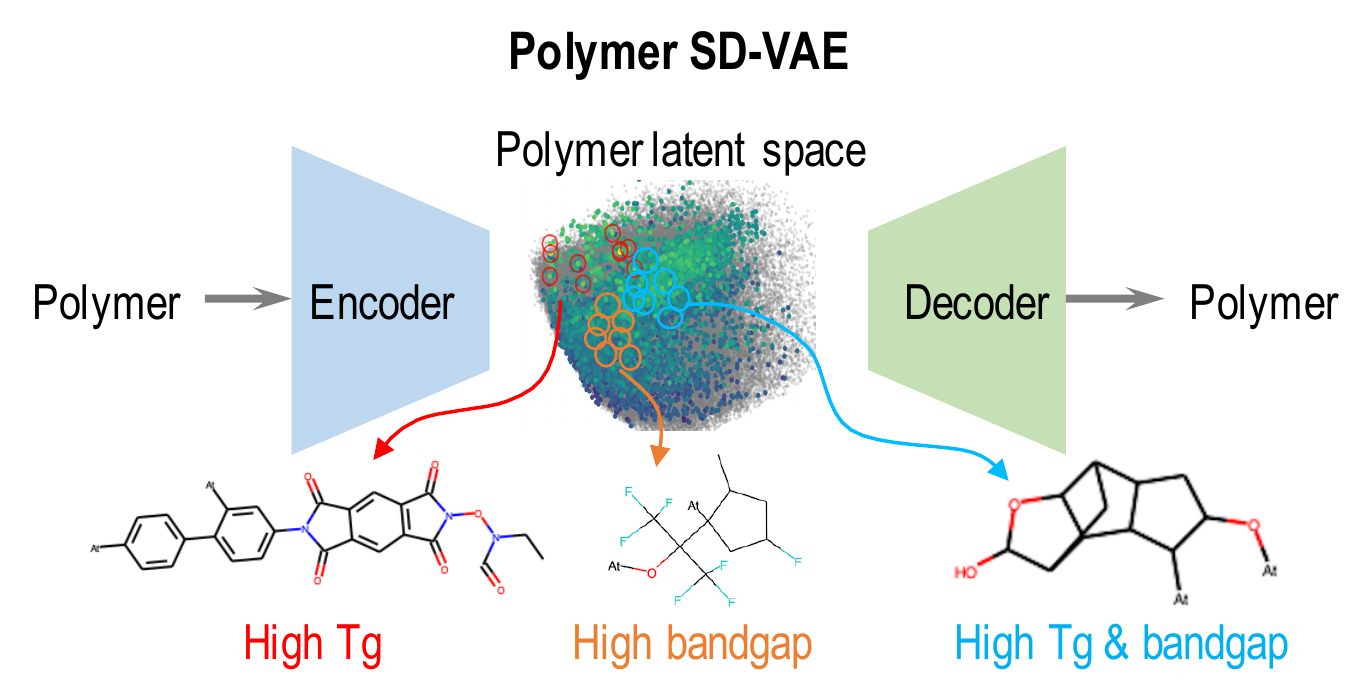}
		\label{fig:TOC}
	\end{center}
\end{figure}

\end{document}